\documentclass[twocolumn, fleqn]{article}

\usepackage[utf8]{inputenc}
\usepackage[T1]{fontenc}
\usepackage[a4paper, margin=0.75in]{geometry}
\usepackage[hidelinks]{hyperref}
\usepackage{amsmath}
\usepackage[list-units=single]{siunitx}
\usepackage[version=4]{mhchem}
\usepackage{tikz}
\usepackage{graphicx}
\usepackage{float}
\usepackage{subcaption}
\usepackage{longtable}
\usepackage{booktabs}
\usepackage{multirow}
\usepackage{bigdelim}
\usepackage{multicol}
\usepackage[capitalize, noabbrev, nameinlink]{cleveref}
\usepackage[style=nature]{biblatex}
\usepackage{titling}
\usepackage{authblk}
\usepackage{xcolor}
\usepackage{xr}

\DeclareUnicodeCharacter{0394}{\ensuremath{\Delta}}

\DeclareSIUnit{\angstrom}{\textup{\smash{Å}\vphantom{A}}}
\DeclareSIUnit{\atom}{\text{atom}}
\DeclareSIUnit{\bar}{\text{bar}}
\DeclareSIUnit{\rydberg}{\text{Ry}}

\hypersetup{
    hypertexnames=false
}

\newcommand{\subfiglabel}[1]{\begin{subfigure}{0em}\phantomsubcaption\label{#1}\end{subfigure}}

\title{How accurate are foundational machine learning interatomic potentials for heterogeneous catalysis?}
\author[1,*]{Luuk H. E. Kempen}
\author[1]{Raffaele Cheula}
\author[1,*]{Mie Andersen}
\affil[1]{Center for Interstellar Catalysis, Department of Physics and Astronomy, Aarhus University, Aarhus C, Denmark}
\affil[*]{Corresponding authors: \url{luuk@phys.au.dk}, \url{mie@phys.au.dk}}

\begin{document}

\twocolumn[
    \begin{@twocolumnfalse}
        \maketitle

        Foundational machine learning interatomic potentials (MLIPs) are being developed at a rapid pace, promising closer and closer approximation to \emph{ab initio} accuracy.
This unlocks the possibility to simulate much larger length and time scales.
However, benchmarks for these MLIPs are usually limited to ordered, crystalline and bulk materials.
Hence, reported performance does not necessarily accurately reflect MLIP performance in real applications such as heterogeneous catalysis.
Here, we systematically analyze zero-shot performance of \num{80} different MLIPs, evaluating tasks typical for heterogeneous catalysis across a range of different data sets, including adsorption and reaction on surfaces of alloyed metals, oxides, and metal-oxide interfacial systems.
We demonstrate that current-generation foundational MLIPs can already perform at high accuracy for applications such as predicting vacancy formation energies of perovskite oxides or zero-point energies of supported nanoclusters.
However, limitations also exist. We find that many MLIPs catastrophically fail when applied to magnetic materials, and structure relaxation in the MLIP generally increases the energy prediction error compared to single-point evaluation of a previously optimized structure.
Comparing low-cost task-specific models to foundational MLIPs, we highlight some core differences between these model approaches and show that---if considering only accuracy---these models can compete with the current generation of best-performing MLIPs.
Furthermore, we show that no single MLIP universally performs best, requiring users to investigate MLIP suitability for their desired application.

    \end{@twocolumnfalse}

    \vspace*{2em}
]

\section{Introduction}

Data-driven molecular and materials modelling has seen enormous progress in the last five years~\cite{kulichenko2024,choi2025,omranpour2025}.
This development has been driven by advances in graph neural networks and other novel architectures~\cite{schutt2021,gasteiger2022a,gasteiger2021,gasteiger2022b,batzner2022,tan2025,musaelian2023,park2024,batatia2023,bochkarev2024,zitnick2022,passaro2023,liao2023,liao2024a,liao2024b,fu2025} as well as the assembly of large data sets with millions of calculated materials or molecules~\cite{chanussot2021,tran2023,barrosoluque2024,kaplan2025,schmidt2023,sahoo2025,sriram2025,gharakhanyan2025,levine2025,allam2025}.
Together, these have enabled foundational machine learning interatomic potentials (MLIPs)---also known as pretrained or universal MLIPs---that can be applied at a very low computational cost across the periodic table \emph{with reasonable, but varying, accuracy}~\cite{deng2023,batatia2025,kim2024,lysogorskiy2025,rhodes2025,wood2025}.

Despite these advances, MLIPs are typically benchmarked on test splits of the same data sets they were trained on.
Such benchmarks rarely reflect their performance in real applications such as heterogeneous catalysis.
For instance, real catalysts may be amorphous (i.e., they lack long-range order) and may dynamically adapt their shape and composition according to the reaction conditions~\cite{goldsmith2017,chee2023}.
In contrast, benchmark data sets usually contain ordered, crystalline materials.
The Matbench Discovery framework, for instance, focuses on evaluating crystal stability predictions~\cite{riebesell2025}.
A further limitation is that MLIPs are often trained to reproduce stable structures and energies.
For heterogeneous catalysis, however, we also require accurate predictions of reaction barriers, i.e., both adsorption and transition state (TS) energies~\cite{schaaf2023,perego2024,yang2025}.
These are then incorporated into microkinetic models to determine reaction mechanisms, catalyst activity, and product selectivity~\cite{andersen2019,motagamwala2021}.
Because rates depend exponentially on barriers, a very high accuracy is required both from the MLIP and from the underlying method used to generate the training data, typically density functional theory (DFT).

\begin{figure*}
    \centering
    \includegraphics{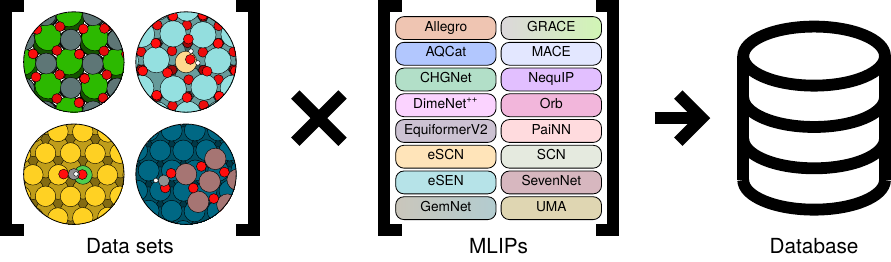}
    \caption{An overview of the data sets and foundational MLIPs assessed in this work.}
    \label{fig:overview}
\end{figure*}

In this work we critically assess the performance of a wide range of recently developed foundational MLIPs for typical tasks in heterogeneous catalysis modelling (\cref{fig:overview}), including the calculation of surface vacancy formation energies, adsorption energies, TS energies, and vibrational properties in the form of zero-point vibrational energies.
We show that both the MLIP architecture as well as the data set it has been trained on are important factors in determining MLIP performance.
Generally, the eSEN, Orb, and UMA models are among the best-performing models across most of the evaluated properties, and no single model stands out as being best across all data sets and properties evaluated in this work.
This highlights the need for exploring MLIP performance for a desired application before choosing a model.
We further highlight that a large number of evaluated MLIPs catastrophically fail when applied to surfaces containing magnetic materials.
We identify a clear correlation between this catastrophic failure and the training data sets utilized to train the various MLIPs.
While most of the comparisons in this work are done using DFT-relaxed geometries as reference to enable screening of a large number of models, we also analyze the effect of MLIP relaxation on a smaller subset of data, showing that this increases the error of the best-performing models.
Finally, we compare the performance of MLIPs to that of low-cost, task-specific models trained on the analyzed data sets, highlighting areas where current-generation foundational potentials are lagging behind these task-specific models when used out-of-the-box (zero-shot), without fine-tuning.

\section{Methods}

\subsection{Data collection}

For this study, a database (\cref{fig:overview}) was built up containing potential energy evaluations of \num{6} data sets with relevance to catalysis (see \cref{sec:methods/datasets}), using \num{80} different pretrained foundational MLIPs (see \cref{sec:methods/models}).
The creation of this database was largely automated by a workflow that evaluates and stores results after specifying a combination of data set, MLIP, and result type (for example, potential energy) to obtain.
These results are stored in a database, after which analysis can be performed by calculating target properties (such as reaction energy) from the relevant results (in case of reaction energy, the potential energies of pairs of initial and final states).
This high-throughput screening is enabled by using the atomic simulation environment (ASE)~\cite{larsen2017} interfaces exposed by the various MLIP packages.

To enable the possibility of screening this large amount of models, the primary focus of this work is on single-point energy evaluations of DFT-relaxed geometries.
This is not a fully realistic use case, though, as a DFT relaxation is computationally expensive and would practically be replaced by MLIP relaxations.
At this scale, however, that also becomes computationally expensive.
Hence, we discuss this more realistic use case only for a smaller subset of the data sets (\cref{sec:results/relaxation}).

\subsection{Data sets}  \label{sec:methods/datasets}

The data sets utilized in this study can broadly be divided into three categories: perovskite oxide surfaces, reactions on alloyed metal and oxide surfaces, and metal--oxide interfaces.
These data sets are summarized in \cref{tab:datasets}.

\subsubsection{Perovskite oxide surfaces}

The perovskite oxide surface data set, obtained from \textcite{taylor2025}, explores the energetics of nine (001) perovskite oxide (\ce{$AB$O3}) surfaces ($A$ = \ce{Ca}, \ce{Sr}, \ce{Ba}; $B$ = \ce{Ti}, \ce{Zr}, \ce{Sn}).
This data set includes \ce{$A$O} and \ce{$B$O2} surface terminations of all perovskite oxides, as well as vacancies of \ce{$A$}, \ce{$B$} and \ce{O} sites at various depths in the slabs.
The target DFT energies in this data set were evaluated using the Vienna \emph{ab initio} simulation package (VASP)~\cite{kresse1996a,kresse1996b}, using the projector augmented-wave (PAW) method~\cite{blochl1994} and a plane-wave basis set energy cutoff of \qty{700}{\electronvolt}.
The exchange--correlation functional was approximated using the Perdew--Burke--Ernzerhof (PBE) functional~\cite{perdew1996}.
More details on the creation of this data set can be found in \textcite{taylor2025}.

The target property evaluated for this data set in this study is the slab vacancy formation energy, which is defined by \textcite{taylor2025} with respect to the corresponding bulk vacancy as
\begin{equation}
    E_{\mathrm{form},X}^i = \left( E_{X\text{-vac}}^i + N_{\mathrm{cells}} E_{\ce{$AB$O3}}^{\mathrm{bulk}} \right) - \left( E_{\mathrm{slab}}^i + E_{X\text{-vac}}^{\mathrm{bulk}} \right),
\end{equation}
where $E_{X\text{-vac}}^i$ is the energy of the slab with site $X$ vacated and surface termination $i$, $E_{\ce{$AB$O3}}^{\mathrm{bulk}}$ is the energy of the bulk perovskite oxide, $E_{\mathrm{slab}}^i$ is the energy of the pristine slab with surface termination $i$, and $E_{X\text{-vac}}^{\mathrm{bulk}}$ is the energy of the bulk perovskite oxide with site $X$ vacated.
$N_{\mathrm{cells}}$ describes how many supercells of the bulk perovskite oxide are included in the vacated bulk perovskite oxide system, which is \num{8} (\numproduct{2x2x2}) for this data set.

\subsubsection{Reactions on alloyed metal and oxide surfaces}

The category of reactions on alloyed metal and oxide surfaces encompasses two data sets.
The first data set, obtained from \textcite{cheula2024}, concerns the \ce{ZrO2(101)} surface, both pristine and doped with \num{16} different metals.
The target DFT energies in this data set were evaluated using Quantum Espresso~\cite{giannozzi2009,giannozzi2017} with the PBE exchange--correlation functional~\cite{perdew1996} and pseudopotentials from the SSSP library~\cite{prandini2018}.
A plane-wave basis set was used with a plane-wave cutoff energy of \qty{40}{\rydberg} (\qty{\sim544}{\electronvolt}) and an electronic density cutoff energy of \qty{320}{\rydberg} (\qty{\sim4354}{\electronvolt}).
A Hubbard $U$ correction of \qty{4}{\electronvolt} was applied to the d orbitals of \ce{Zr} atoms~\cite{anisimov1991,cococcioni2005}.
Elementary reaction steps for \ce{CO2} hydrogenation to methanol were investigated on the pristine \ce{ZrO2} surface as well as \num{8} metal-doped surfaces.
More details on the creation of this data set can be found in \textcite{cheula2024}.

Four target properties are evaluated for the data set in this study.
The first two are the activation energy $E_{\mathrm{act}}$ and reaction energy $\Delta E_{\mathrm{r}}$ for elementary reaction steps.
The final two are the formation energies for adsorbates and TSs with respect to the clean slabs and gas-phase \ce{H2}, \ce{H2O}, and \ce{CO2} following the definition by \textcite{cheula2024}:
\begin{equation}
    E_{\mathrm{form}} = E_{\mathrm{slab + ads}} - E_{\mathrm{slab}} - N_{\ce{C}} E_{\ce{C}} - N_{\ce{O}} E_{\ce{O}} - N_{\ce{H}} E_{\ce{H}},  \label{eq:formation-energy}
\end{equation}
where $E_{\mathrm{slab + ads}}$ is the energy of a slab with adsorbate or TS, $E_{\mathrm{slab}}$ is the energy of a clean slab, $N_i$ are the number of atoms of species $i$ in the adsorbate, and $E_i$ are reference energies of these atoms.
These reference energies are calculated with \ce{H2}, \ce{H2O}, and \ce{CO2} as reference, such that
\begin{align}
    E_{\ce{H}} &= \frac{1}{2} E_{\ce{H2}};  \label{eq:ref-h} \\
    E_{\ce{O}} &= E_{\ce{H2O}} - E_{\ce{H2}}; \\
    E_{\ce{C}} &= E_{\ce{CO2}} - 2 E_{\ce{O}}.
\end{align}

The second data set, obtained from \textcite{cheula2025}, encompasses (111) and (100) facets of \num{6} pure metals and \num{12} single-atom alloys (SAAs), including elementary reaction steps for the reverse water--gas shift (RWGS) reaction.
The target DFT energies in this data set were evaluated using the same DFT settings as in the (doped) \ce{ZrO2(101)} data set outlined above, with the exception of using the BEEF-vdW exchange--correlation functional~\cite{wellendorff2012} instead of PBE and without Hubbard $U$ correction.
The same target properties were evaluated as for the (doped) \ce{ZrO2(101)} data set: activation energy, reaction energy, and adsorbate and TS formation energy.
However, in this case, analogous to the earlier work, the formation energies are defined with respect to gas-phase \ce{H2}, \ce{H2O}, and \ce{CO}, such that in this case
\begin{equation}
    E_{\ce{C}} = E_{\ce{CO}} - E_{\ce{O}}
\end{equation}
is substituted into \cref{eq:formation-energy}.

\subsubsection{Metal--oxide interfaces}  \label{sec:methods/datasets/metal-oxide}

The category of metal--oxide interfaces encompasses three closely-related data sets.

The first data set, obtained from \textcite{kempen2025a}, includes global minima of metal oxide (\ce{Zn_{y}O_{x}} and \ce{In_{y}O_{x}}) nanoclusters on metal (\ce{Cu}, \ce{Pd}, and \ce{Au}) surfaces.
The target DFT energies in this data set were evaluated using the GPAW package~\cite{mortensen2005,enkovaara2010,mortensen2024}, which implements the PAW method~\cite{blochl1994}, with the PBE exchange--correlation functional~\cite{perdew1996}.
The wave functions were represented using a plane-wave basis with a plane-wave energy cutoff of \qty{400}{\electronvolt} for \ce{Cu}- and \ce{Au}-supported systems, and \qty{500}{\electronvolt} for \ce{Pd}-supported systems.
A \numproduct{2x2x1} Monkhorst--Pack $k$-point grid~\cite{monkhorst1976} was utilized for Brillouin zone sampling.
More details on the creation of this data set can be found in \textcite{kempen2025a}.

The target property evaluated for this data set is the zero-point energy of the nanoclusters, used by \textcite{kempen2025a} to perform a free energy analysis.
The zero-point energy is evaluated in the harmonic limit, considering only the vibrational modes of the nanocluster atoms.
To evaluate vibrational modes, the nanoclusters were relaxed to a force criterion of $F_{\mathrm{max}} = \qty{0.01}{\electronvolt\per\angstrom}$ after which the cluster atoms were individually displaced in the positive and negative direction of the three Cartesian coordinate axes with a step size of \qty{0.01}{\angstrom} to calculate a second-order central finite difference approximation of the Hessian matrix, as implemented in the ASE vibrations module~\cite{larsen2017}.
The zero-point energy is then given as
\begin{equation}
    E_{\mathrm{ZPE}} = \frac{1}{2} \sum_{i=1}^{3N} \varepsilon_i,  \label{eq:zpe}
\end{equation}
where $\varepsilon_i$ is the vibrational energy of mode $i$.

The second data set, obtained from \textcite{nielsen2025}, includes a large range of formate (\ce{HCOO}) adsorbates on the metal-supported metal oxide nanoclusters described above, including both global minima and local minima of the nanoclusters within \qty{0.1}{\electronvolt} of the global minimum energy structure.
The target DFT energies in this data set were evaluated with the GPAW package~\cite{mortensen2005,enkovaara2010,mortensen2024} and the PBE exchange--correlation functional~\cite{perdew1996} with D4 dispersion corrections~\cite{caldeweyher2017,caldeweyher2019,caldeweyher2020}.
A plane-wave energy cutoff of \qty{400}{\electronvolt} was used for all systems, as well as a \numproduct{2x2x1} Monkhorst--Pack $k$-point grid~\cite{monkhorst1976}.
More details on the creation of this data set can be found in \textcite{nielsen2025}.

The target property evaluated for the data set in this study is the adsorption energy of the formate adsorbate, defined as
\begin{equation}
    E_{\mathrm{ads}} = E_{\mathrm{cluster + ads}} - E_{\mathrm{cluster}} - E_{\ce{HCOO (g)}},
\end{equation}
where $E_{\mathrm{cluster + ads}}$ is the energy of a nanocluster with the adsorbate, $E_{\mathrm{cluster}}$ is the energy of the clean nanocluster, and $E_{\ce{HCOO (g)}}$ is the energy of gas-phase formate.

The third data set, obtained from \textcite{kempen2025b}, includes initial, final, and TSs for the \ce{CO2 + H <--> HCOO} elementary reaction step on \ce{In_{y}O_{x}/Cu(111)} inverse catalyst systems.
The target DFT energies in this data set were evaluated using the GPAW package~\cite{mortensen2005,enkovaara2010,mortensen2024} and the PBE exchange--correlation functional~\cite{perdew1996} with D4 dispersion corrections~\cite{caldeweyher2017,caldeweyher2019,caldeweyher2020}.
A plane-wave energy cutoff of \qty{600}{\electronvolt} was used for all systems, as well as a \numproduct{2x2x1} Monkhorst--Pack $k$-point grid~\cite{monkhorst1976}.
More details on the creation of this data set can be found in \textcite{kempen2025b}.

The target properties evaluated for this data set in this study are the activation energy $E_{\mathrm{act}}$ and reaction energy $\Delta E_{\mathrm{r}}$ for the \ce{CO2 + H <--> HCOO} reaction step on the different active sites.

\begin{table*}
    \centering
    \caption{Overview of the data sets utilized in this study.}
    \label{tab:datasets}
    \begin{tabular}{lll}
        \toprule
        Data set & Ref. & Target properties (number of entries) \\
        \midrule
        Perovskite oxide surfaces & \cite{taylor2025} & Slab vacancy formation energy (\num{305}) \\
        \midrule
        \multirow{4}{*}{Reactions on (doped) zirconia} & \multirow{4}{*}{\cite{cheula2024}} & Activation energy (\num{62}) \\
        && Reaction energy (\num{62}) \\
        && Adsorbate formation energy (\num{122}) \\
        && TS formation energy (\num{62}) \\
        \midrule
        \multirow{4}{*}{Reactions on metals and SAAs} & \multirow{4}{*}{\cite{cheula2025}} & Activation energy (\num{632}) \\
        && Reaction energy (\num{632}) \\
        && Adsorbate formation energy (\num{1481}) \\
        && TS formation energy (\num{632}) \\
        \midrule
        Metal oxide nanoclusters & \cite{kempen2025a} & Zero-point energy (\num{180}) \\
        \midrule
        Formate on metal oxide nanoclusters & \cite{nielsen2025} & Adsorption energy (\num{2869}) \\
        \midrule
        \multirow{2}{*}{Formate formation on metal oxide nanoclusters} & \multirow{2}{*}{\cite{kempen2025b}} & Activation energy (\num{144}) \\
        && Reaction energy (\num{144}) \\
        \bottomrule
    \end{tabular}
\end{table*}

\subsection{Machine learning interatomic potentials}  \label{sec:methods/models}

In this study, \num{80} different pretrained MLIPs have been utilized, spanning \num{16} different families---groups of models with broadly the same architecture.
A short description of each of these families, and the pretrained models within each architecture, is given below.
We further specify which training data sets the pretrained models have been trained on. Here, we denote combination of data sets A and B as `A\,+\,B,' and pre-training on data set A followed by fine-tuning on data set B as `A\,→\,B.'
A full overview of all models included in this study is given in \cref{supp/tab:mlips}.

CHGNet~\cite{deng2023} is a GNN architecture where magnetic moments are explicitly included.
We evaluated CHGNet performance with the model trained on the MPTrj data set~\cite{deng2023}.

DimeNet\textsuperscript{++}~\cite{gasteiger2022a} is a GNN architecture where messages between atoms are embedded in addition to atomic features.
We evaluated DimeNet\textsuperscript{++} performance with the model trained on OC20-All~\cite{chanussot2021}.

GemNet~\cite{gasteiger2021} is based on DimeNet\textsuperscript{++}, but utilizes geometric message passing and is equivariant.
We evaluated GemNet-dT performance with the models trained on OC20-All~\cite{chanussot2021}, as well as OC22~\cite{tran2023}.
GemNet-OC~\cite{gasteiger2022b} is a further development of GemNet, specifically built for the OC20 data set.
We evaluated GemNet-OC performance with the models trained on OC20-All~\cite{chanussot2021} as well as on OC22, OC20-All\,+\,OC22, and OC20\,→\,OC22~\cite{tran2023}.

The PaiNN architecture~\cite{schutt2021} is another example of an equivariant GNN.
Here, we evaluated PaiNN performance with the model trained on OC20-All~\cite{chanussot2021}.

The NequIP network architecture~\cite{batzner2022,tan2025} is an equivariant GNN architecture.
NequIP MLIPs with $\ell_{\mathrm{max}} = 3$ (maximum spherical harmonic rotation order included in the basis set) and \num{6} message-passing layers are available, trained on both MPTrj~\cite{deng2023} and OMat24\,→\,MPTrj\,+\,sAlex (OAM)~\cite{barrosoluque2024}.

Similarly, the Allegro network architecture~\cite{musaelian2023,tan2025} is an equivariant graph neural network architecture.
Allegro MLIPs with $\ell_{\mathrm{max}} = 3$ and \num{4} message-passing layers are available, trained on both MPTrj~\cite{deng2023} and OAM~\cite{barrosoluque2024}.

SevenNet~\cite{park2024} is based on the NequIP architecture, introducing a parallelization scheme for scaling to larger systems via spatial decomposition.
While this feature is not relevant for the application in this study, we still evaluated the four currently available pretrained SevenNet models.
These include two models trained on MPTrj~\cite{deng2023} (one larger model with $l_{\mathrm{max}} = 3$), one model trained on OMat24~\cite{barrosoluque2024}, and one model trained simultaneously on MPTrj, OMat24, and sAlex~\cite{barrosoluque2024} using multi-fidelity learning~\cite{kim2024}.
Performance of this latter model was evaluated for both MPTrj\,+\,sAlex and OMat24 tasks.

The MACE architecture~\cite{batatia2023} combines equivariant message passing with atomic cluster expansion (ACE).
The MACE foundation models~\cite{batatia2025} pretrained on MPTrj~\cite{deng2023} (MP-0a in three sizes and MP-0b3), MPTrj\,+\,sAlex~\cite{deng2023,barrosoluque2024}, OMat24~\cite{barrosoluque2024}, and MatPES-PBE~\cite{kaplan2025} were used to evaluate MACE model performance.

Similarly, the GRACE architecture~\cite{bochkarev2024} is an extension of ACE, incorporating graphs.
GRACE foundation models~\cite{lysogorskiy2025} with one- and two-layer architectures, three sizes, and trained on MPTrj~\cite{deng2023}, OMat24~\cite{barrosoluque2024}, and OAM~\cite{barrosoluque2024} have been used to evaluate GRACE model performance.

The Orb family of models~\cite{rhodes2025} is trained using diffusion pretraining with subsequent fine-tuning.
Their architecture is not equivariant by default, but equivariance is approximated using equigrad, a regularization scheme.
All eight publicly-released Orb models---combinations of the three key variables conservatism (direct or conservative), maximum neighbor limits (20 or infinity), and training data set (MPTrj\,+\,Alex~\cite{deng2023,schmidt2023} or OMat24~\cite{barrosoluque2024})---were used to evaluate Orb model performance.

The SCN architecture~\cite{zitnick2022} is a GNN architecture utilizing spherical functions.
We evaluated SCN performance with the models trained on OC20-2M and OC20-All+MD~\cite{chanussot2021}.

The eSCN architecture~\cite{passaro2023} builds on SCN by enforcing equivariance in message passing.
We evaluated eSCN performance with the two models trained on OC20-All+MD~\cite{chanussot2021}: one with $M = 2$ and \num{12} layers, and one with $M = 3$ and \num{20} layers.

The EquiformerV2 architecture~\cite{liao2024a} builds on Equiformer~\cite{liao2023}, an equivariant GNN, by introducing eSCN convolutions, among other tweaks.
We evaluated EquiformerV2 performance with the 31M, 86M (83M for OC20), and 153M models trained on MPTrj~\cite{deng2023}, OMat24~\cite{barrosoluque2024}, OAM~\cite{barrosoluque2024}, and OC20 (OC20-All+MD for 31M and 153M, and OC20-2M for 83M)~\cite{chanussot2021}.
We additionally evaluated performance with the DeNS models~\cite{liao2024b} trained on MPTrj, which introduce an auxiliary task to denoise non-equilibrium structures, and with the publicly-available model trained on OC22 ($\lambda_E = 4$, $\lambda_F = 100$), which has 121M parameters~\cite{tran2023}.

The eSEN architecture~\cite{fu2025} builds on eSCN and Equiformer, with a number of changes to the internal architecture of the convolution blocks to obtain smooth models.
We evaluated eSEN performance with the 30M models~\cite{fu2025} trained on MPTrj~\cite{deng2023}, OMat24~\cite{barrosoluque2024}, and OAM~\cite{barrosoluque2024}, as well as with the publicly available small and medium models trained on OC25 (eSEN-S-cons.~and eSEN-M-d.)~\cite{sahoo2025}.

The UMA architecture~\cite{wood2025} builds on eSEN, expanding its input to incorporate a global embedding based on total charge, spin multiplicity, and the DFT task.
Incorporation of the DFT task allows the model to be trained on a mix of DFT tasks.
Furthermore, UMA introduces a mixture of linear experts to increase the model size while keeping inference times manageable.
UMA models have been trained on a combination of extended OC20~\cite{chanussot2021,wood2025}, ODAC25~\cite{sriram2025}, OMat24~\cite{barrosoluque2024}, OMC25~\cite{gharakhanyan2025}, and OMol25~\cite{levine2025} data sets.
In this study, we evaluated UMA performance for the OC20 and OMat24 tasks, using the small- and medium-size models, version 1.1 (UMA-S-1.1 and UMA-M-1.1).
Furthermore, we also evaluated UMA performance for a UMA-M-1.1 model trained on OAM~\cite{barrosoluque2024}.

The AQCat model family~\cite{allam2025} contains models based on EquiformerV2, introducing additive shifts at various points in the model using feature-wise linear modulation (FiLM)~\cite{perez2018}.
These shifts incorporate task information on spin polarization treatment and fidelity.
These models are trained on a combination of OC20~\cite{chanussot2021} and the AQCat25 data set~\cite{allam2025}, which is an expansion of OC20 with explicit inclusion of spin polarization, which is used as a training feature.
In this study, we evaluated AQCat performance using the four publicly-available models.

The training data sets utilized to train the models discussed above were not all created equally; most importantly, they do not all use the same exchange--correlation functional.
The MPTrj~\cite{deng2023}, OC22~\cite{tran2023}, and OMat24~\cite{barrosoluque2024} data sets consist of a mixture of PBE~\cite{perdew1996} and $\mathrm{PBE}+U$ calculations, following Materials Project recommendations~\cite{jain2013}.
On the other hand, the Alex~\cite{schmidt2023} (and thus sAlex~\cite{barrosoluque2024}) and MatPES-PBE~\cite{kaplan2025} data sets include only PBE calculations.
The OC20~\cite{chanussot2021}, OC25~\cite{sahoo2025}, and AQCat25~\cite{allam2025} data sets utilize the RPBE exchange--correlation functional~\cite{hammer1999}; the OC25 data set further applies a D3 correction with zero damping~\cite{grimme2010}.

Because of the complexity of visualizing all evaluated MLIPs in figures, the MLIPs were divided into three dimensions for visualization: model family, indicated with color; training data set, indicated with markers; and model size (categorized into small, medium, and large), indicated with marker size.
Not all MLIPs fit neatly within these dimensions, though, and some groups of distinct MLIPs have identical symbols under the scheme outlined above.
Therefore, for these groups, only the best-performing MLIP (ranked by RMSE) is shown in the figures.
These groups are as follows: GemNet-OC with and without strict neighbor enforcement; SevenNet-0 and SevenNet-l3i5 trained on MPTrj; MACE-MP-0a (three sizes) and MACE-MP-0b3; Orb models with the same training data sets; and EquiformerV2-31M trained on MPTrj (with and without DeNS).
The error values for all MLIPs, including the ones that are not shown in the figures, are available in the tables in \cref{supp/sec:model-performance}.

\subsection{Gas-phase error correction}  \label{sec:methods/gas-phase-correction}

Most target properties described above are calculated from purely solid-state geometries: bulks or slabs, possibly with adsorbates on top of these slabs.
For example, the activation energy is the difference between the energy of the TS and the initial state of the reactants, adsorbed to the surface in both cases.
Two exceptions exist to this: the formation energy (both for adsorbate and TS) and the adsorption energy.
Both are calculated from a combination of solid-state and gas-phase geometries.
However, all MLIPs investigated in this work target prediction of solid-state geometries and are trained with data sets containing only those geometries.
Therefore, it cannot be expected that these MLIPs predict energies of gas-phase molecules with high accuracy.
For a more fair comparison of MLIP accuracy, we present errors for the prediction of these properties after correcting the MLIP property predictions by the mean error across the entire set of predicted properties (except when indicated otherwise):
\begin{equation}
    E_{\mathrm{MLIP},i}^{\mathrm{corr}} = E_{\mathrm{MLIP},i} - \frac{1}{N} \sum_j ( E_{\mathrm{MLIP},j} - E_{\mathrm{DFT},j} ),
\end{equation}
where $E_{\mathrm{MLIP},i}$ is the MLIP-predicted energy value for property $i$, $E_{\mathrm{DFT},i}$ is the DFT target value for property $i$, and $N$ is the total number of property values.
This offset is assumed to mostly comprise the error due to poor predictions of gas-phase geometries.
Note that the correction value is calculated over the entire data set, not split by molecule, as some molecules exist in very few data points and thus the error correction might not accurately reflect gas-phase prediction error.

In addition, the GemNet-dT model architecture cannot predict properties for systems with two atoms or less, and thus formation energies defined using \cref{eq:formation-energy} (relying on the energy prediction of \ce{H2} through \cref{eq:ref-h}) cannot be calculated with this architecture.

\section{Results}

\subsection{Model performance trends}  \label{sec:results/trends}

We begin by comparing the zero-shot performance of the different MLIPs for a subset of data sets and target properties, picking a number of interesting observations.
This performance is quantified through the root mean square error (RMSE) and maximum absolute error (MaxAE) between the MLIP predictions and DFT energies.
We again highlight that these results are a comparison of DFT-relaxed geometries, except when indicated otherwise.
Furthermore, note that the DFT functional employed in the data sets used to train the foundational MLIPs (\cref{sec:methods/models}) may not match the functional used in the target property evaluation (\cref{sec:methods/datasets}).
It is therefore possible that small changes to the ranking of the foundational MLIPs would occur if the target property were to be reevaluated with a different functional.
Prediction performance for all MLIPs and all target properties is given in \cref{supp/sec:model-performance}.

\begin{figure*}
    \centering
    \includegraphics{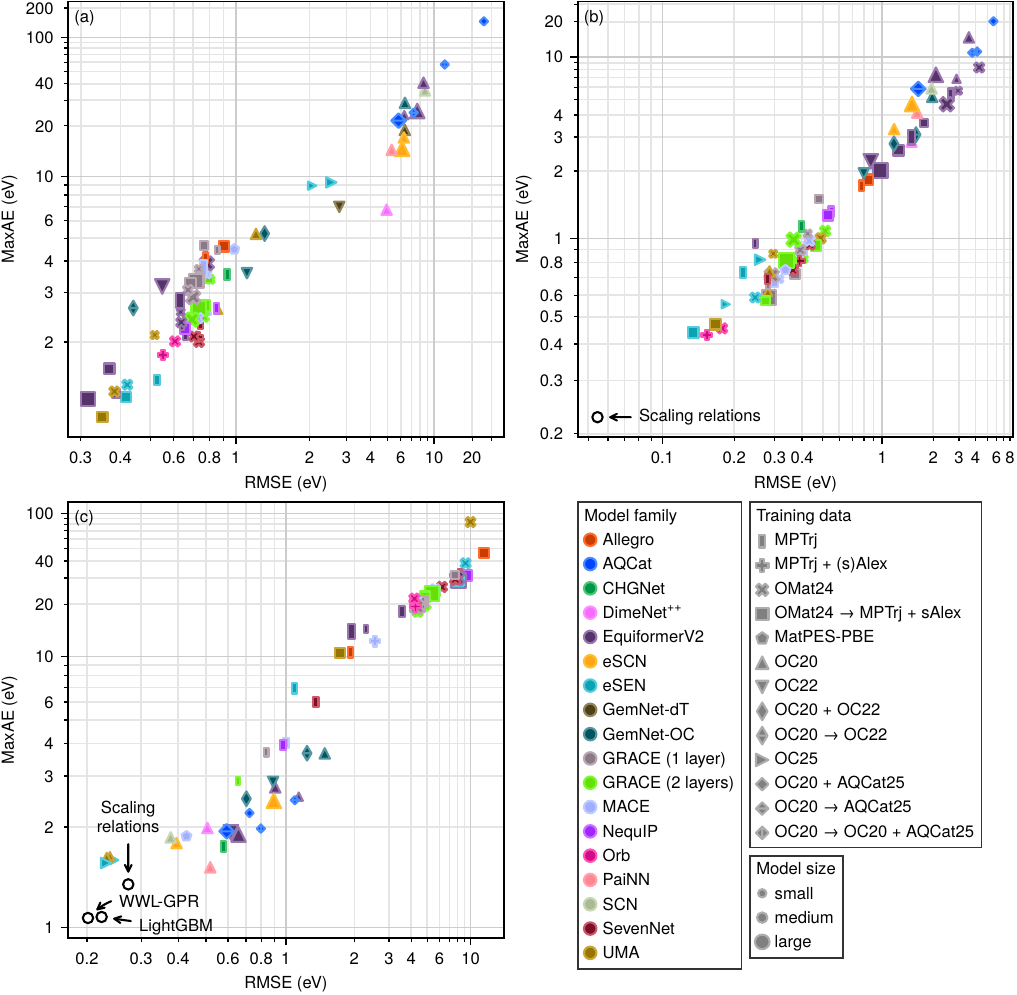}
    \caption{Prediction performance of the MLIPs for three different data sets and target properties: (a) slab vacancy formation energy of perovskite oxide surfaces; (b) formation energy of TSs of \ce{CO2} hydrogenation reaction steps on (doped) \ce{ZrO2(101)} surfaces; (c) formation energy of TSs of RWGS reaction steps on metals and SAAs. Larger versions of these figures and data are available in \cref{supp/sec:model-performance}.}    \subfiglabel{fig:perovskites-slab-vacancy-formation-energy}    \subfiglabel{fig:zirconia-formation-energy-ts}    \subfiglabel{fig:single-atom-alloys-formation-energy-ts}    \label{fig:prediction-combinations}
\end{figure*}

In \cref{fig:perovskites-slab-vacancy-formation-energy}, we show MLIP prediction performance for the slab vacancy formation energy of perovskite oxide surfaces.
These results show that generally, the EquiformerV2, UMA, eSEN, and Orb model architectures are the most well-performing.
However, a stronger trend is visible when analyzing model performance by training data sets.
Models trained on OAM and OMat24 perform best, whereas models trained on OC20 clearly perform worst.
This is unsurprising, given that the OMat24 data set includes inorganic bulk materials including many oxides, and thereby most closely approximates the perovskite oxide data set.
On the other hand, OC20 does not feature oxide structures and focuses on adsorbates, which do not occur in this perovskite oxide data set.
Some of the models trained on OC22, specifically aimed at covering metal oxide materials, perform reasonably well, but do not outperform those trained on OAM or OMat24.

This trend is much less obvious when analyzing MLIP performance for the formation energy of TSs of \ce{CO2} hydrogenation reaction steps on (doped) \ce{ZrO2(101)} surfaces, as shown in \cref{fig:zirconia-formation-energy-ts}.
A large spread can be observed for models trained on training data sets such as OAM and OMat24.
Analogous to the previous example, eSEN, Orb and UMA are the best-performing models, while in this case the EquiformerV2 models are among the worst-performing models.

Trends are again different when analyzing the formation energy of TSs of RWGS reaction steps on metals and SAAs, as shown in \cref{fig:single-atom-alloys-formation-energy-ts}.
Here, the models trained on the OC20, OC22, OC25, and even MPTrj data sets strongly outperform those trained on OAM and OMat24.
A strong contributor here is that there is a large cluster of models---including models trained on OAM and OMat24---in the top right area of \cref{fig:single-atom-alloys-formation-energy-ts}, showing very large prediction errors.
This trend seems to be caused by poor prediction of magnetic elements, and is explored in more detail in the next section.

Overall, the training data set strongly affects model performance, possibly moreso than the model architecture.
The presence of oxide surfaces in the training data set seems to be more important in determining model performance than the presence of adsorbates.
In fact, the zirconia data set shows that models trained on data sets not explicitly including adsorbates can still be highly performant for predicting TS energies of surface reactions.
In general, we see that the eSEN, Orb, and UMA models (when combined with a suitable training data set) are the best-performing models from the set of models evaluated in this work.

\subsection{Poor predictions of magnetic elements}  \label{sec:results/magnetic}

One spurious observation is that a large number of MLIPs catastrophically fail to correctly predict formation energies of both reaction intermediates (\cref{supp/fig:single-atom-alloys-formation-energy-ads}) and TSs (\cref{fig:single-atom-alloys-formation-energy-ts}) on a large number of metal and SAA surfaces.
As shown in these figures, many MLIPs exhibit RMSE values exceeding \qty{3}{\electronvolt}, but more importantly maximum errors exceeding \qty{15}{\electronvolt}, approaching \qty{100}{\electronvolt} for the most extreme cases.

\begin{figure*}
    \centering
    \includegraphics{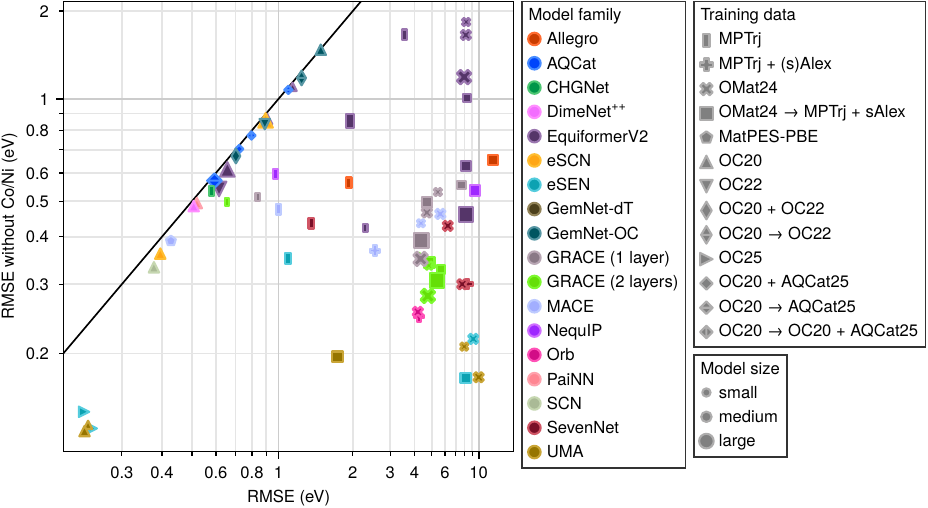}
    \caption{Comparison between RMSEs of formation energy predictions for TSs on metals and SAAs for the entire data set and for the data set with \ce{Co}- and \ce{Ni}-containing surfaces removed. The black line is a parity line. Data is available in \cref{supp/sec:model-performance}.}
    \label{fig:single-atom-alloys-formation-energy-ts-filter-comparison}
\end{figure*}

We further investigated these spurious errors by analyzing formation energy predictions split by surface and adsorbate (for reaction intermediates) and split by surface and reaction (for TSs).
Some representative examples from this analysis are shown in \cref{supp/fig:saa-errors-ads,supp/fig:saa-errors-ts}.
Specifically, the combinations of \ce{C}- and \ce{O}-containing adsorbates on \ce{Co}- and \ce{Ni}-containing surfaces show large errors.
In \cref{fig:single-atom-alloys-formation-energy-ts-filter-comparison}, we show a comparison between RMSEs for the full data set and for the data set with \ce{Co}- and \ce{Ni}-containing surfaces removed.
A subset of models clearly lies on or close to the parity line---indicating the \ce{Co}- and \ce{Ni}-containing surfaces do not significantly impact overall model performance---while many models show much larger RMSE values when \ce{Co}- and \ce{Ni}-containing surfaces are included.
More specifically, the training data set-based trend discussed in the previous section is clearly visible in this figure.
Models trained on OC20, OC22, and OC25 are most performant on \ce{Co}- and \ce{Ni}-containing surfaces, followed by models trained on MPTrj, and finally models trained on OMat24, Alex, or OAM show the largest errors.
We hypothesize that this is caused by a combination of these materials being magnetic materials and the utilization of spin-polarization in the preparation of the training data sets.
The CHGNet model shows the smallest discrepancy from the models trained on MPTrj; this could be due to its explicit inclusion of magnetic moments in training the model.
Furthermore, the AQCat models are very close to the parity line, which might be due to its similar inclusion of spin polarization as an explicit feature at training and inference time.
However, these models are trained on OC20 and derived data; it would be interesting to see how such a model architecture performs when trained on data derived from OMat24.

\subsection{Vibrational modes}

Calculating vibrational modes and their energies, via the Hessian, is crucial for determining free energies, for instance in the harmonic limit.
However, approximating the Hessian using a finite difference approach requires $6N$ additional single-point calculations, where $N$ is the number of atoms for which vibrational modes are to be determined.
Therefore, it is common to evaluate vibrational modes for a subset of representative systems and use these to estimate modes for other systems~\cite{reichenbach2019,cheula2024,cheula2025}.

\begin{figure*}
    \centering
    \includegraphics{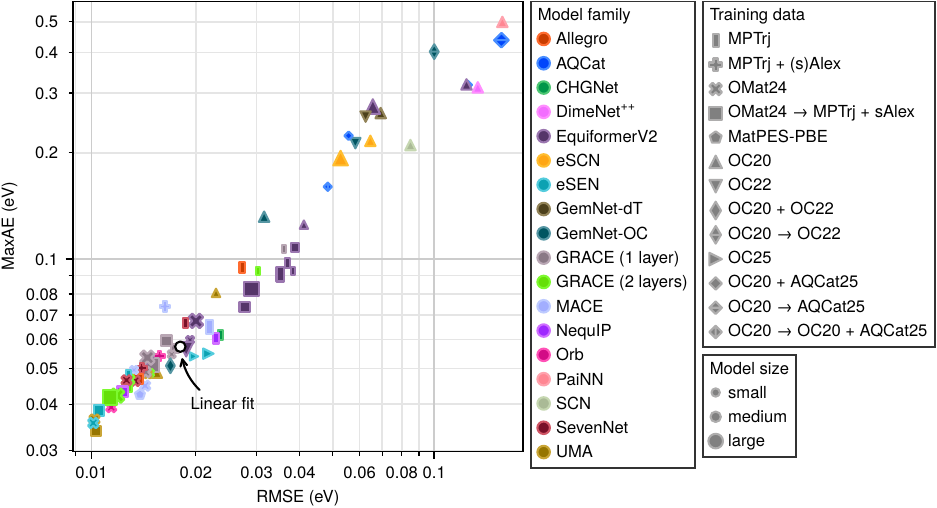}
    \caption{Prediction performance of the MLIPs for the zero-point energy of metal oxide nanoclusters on metal supports. The white circle indicates the linear fitting approach outlined by \textcite{reichenbach2019}. Data is available in \cref{supp/sec:model-performance}.}
    \label{fig:nanoclusters-zero-point-energy}
\end{figure*}

Here, we analyzed MLIP performance for predicting vibrational modes using the finite difference approach, targeting the zero-point energy of metal-supported metal oxide nanoclusters as defined in \cref{eq:zpe}.
To obtain accurate vibrational modes, MLIP relaxations were performed as discussed in \cref{sec:methods/datasets/metal-oxide}.
We also compare MLIP predictions to the linear fitting approach outlined by \textcite{reichenbach2019} and applied to a similar data set.

The resulting comparison (\cref{fig:nanoclusters-zero-point-energy}) shows that the majority of models can predict the zero-point energy with high accuracy, with approximately two-thirds of the evaluated models outperforming the linear fitting approach.
This is in line with earlier work, showing good in-domain performance for the prediction of numerical Hessians using foundational MLIPs~\cite{wander2025}.
It should be noted again that the choice of training data set is correlated with model performance, with models trained on data sets including inorganic materials (OMat24, OAM) generally outperforming models trained on catalysis-focused data sets (OC20, OC22), which are less closely related to the metal-supported metal oxide nanocluster data set.
The eSEN and UMA models are the best performers, reaching RMSEs as low as \qty{0.01}{\electronvolt} compared to the DFT-evaluated zero-point energies.

\subsection{Relaxation in MLIPs}  \label{sec:results/relaxation}

\begin{figure*}
    \centering
    \includegraphics{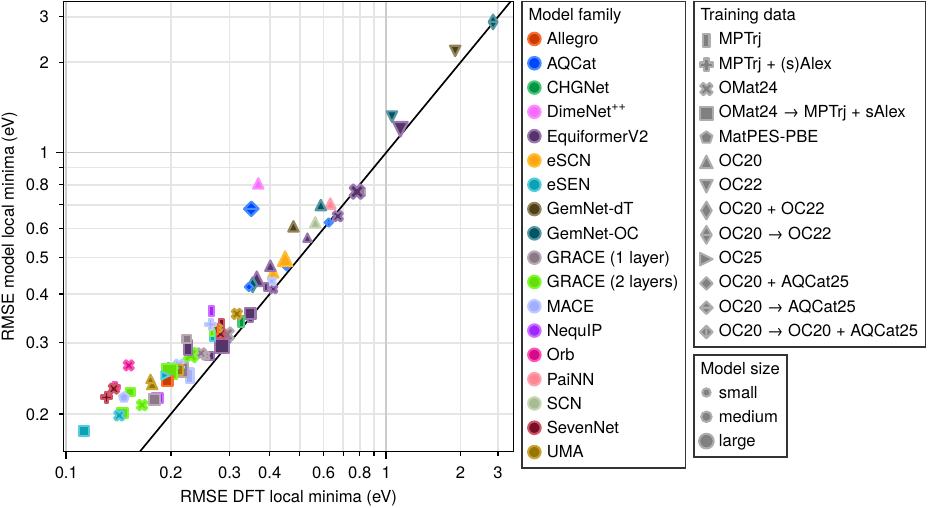}
    \caption{Comparison between RMSEs of adsorption energy predictions for formate on metal-supported oxide nanoclusters when evaluating the MLIP energy of the DFT local minimum and when relaxing into the MLIP local minimum and evaluating the energy of that structure. The black line is a parity line. Data is available in \cref{supp/sec:model-performance}.}
    \label{fig:nanoclusters-formate-relaxed-v-unrelaxed}
\end{figure*}

All comparisons between DFT results and MLIPs in \cref{sec:results/trends,sec:results/magnetic} have been performed by evaluating the energies of DFT-relaxed geometries.
Realistic applications of MLIPs, however, would utilize the MLIP to perform relaxation.
To explore this more realistic use case, we randomly sampled \qty{10}{\percent} of adsorption energies from the formate on metal oxide nanocluster data set, relaxed the corresponding adsorbed structures to a local minimum in each MLIP, and evaluated the resulting energy deviation from the DFT local minimum energy.
The prediction performance for these relaxed structures is shown in \cref{supp/fig:nanoclusters-formate-adsorption-energy-relaxed}.
A comparison between evaluations on the DFT local minima and evaluations on the MLIP local minima is shown in \cref{fig:nanoclusters-formate-relaxed-v-unrelaxed}.
Generally, all models experience an increase in prediction error after relaxing into the MLIP local minima.
For the best-performing models, this increase can be significant, with for instance the eSEN-30M model trained on OAM experiencing a 1.6\texttimes{} increase in RMSE.
More generally, a plateau is observed where RMSEs on the MLIP local minima do not reach below \qty{0.18}{\electronvolt}, despite many models being capable of reaching these RMSE values on the DFT local minima.

\subsection{Comparison to task-specific models}

In contrast to foundational MLIPs, which target application across a wide range of data sets, task-specific models target much more specific applications, usually limited to the (much smaller) data set they have been trained on.
In their respective papers, the data sets for reactions on alloyed metal and oxide surfaces and formate on metal--oxide interfaces have been used to train a range of task-specific models~\cite{cheula2024,cheula2025,nielsen2025}, targeting low-cost, direct prediction of formation energies of adsorbates and TSs.
In this section, we compare performance from these models with that of the foundational MLIPs, giving an indication of the out-of-the-box quality of these MLIPs for catalyst applications compared to task-specific models.

It is important to note that the task-specific models discussed in this section are trained on the same DFT data sets on which their performance is evaluated.
In contrast, the errors reported for the MLIPs are obtained using the models out-of-the-box, without any fine-tuning to the DFT datasets considered here (correcting only for the gas-phase reference when applicable as described in \cref{sec:methods/gas-phase-correction}).

The first set of task-specific models we analyze is the linear scaling relations derived by \textcite{cheula2024} for the formation energy of TSs of \ce{CO2} hydrogenation reaction steps on (doped) \ce{ZrO2(101)} surfaces.
These relations are linear fits of the formation energies of TSs to those of related reaction intermediates.
In \cref{fig:zirconia-formation-energy-ts}, we highlight their accuracy with a white circle. The error reported is a training error. However, owing to the simplicity of the linear fit, test errors are typically rather similar to training errors.
The figure shows that these scaling relations are substantially more accurate than the foundational MLIPs; approximately \num{3} times as accurate as the best-performing model.
However, it should be noted that scaling relations are very limited in their applicability, as a different relation is used to describe each reaction step.

Next, we analyze the set of models derived by \textcite{cheula2025} for the formation energies of both adsorbates and TSs on metals and SAAs.
These models include scaling relations, ensemble decision tree models (LightGBM), and graph kernel-based Gaussian Process Regression (WWL-GPR) models.
We highlight the errors of these models (determined via \num{5}-fold cross-validation) in \cref{supp/fig:single-atom-alloys-formation-energy-ads} for adsorbates and in \cref{fig:single-atom-alloys-formation-energy-ts} for TSs.
Differences between the task-specific models are discussed in more detail by \textcite{cheula2025}, so they are not discussed further here.
The figures show that for adsorbates, the task-specific models show better or similar performance as that of the UMA MLIPs, which are the best-performing MLIPs for this property.
For TSs, the task-specific models (with the exception of the simplest model, the scaling relations) show better performance than the eSEN and UMA MLIPs, especially regarding the maximum error predicted by the models. Here it is also important to note that---similar to the observation for the formate on metal oxide nanocluster data set in \cref{fig:nanoclusters-formate-relaxed-v-unrelaxed}---relaxation in the MLIP may increase the error of the best-performing MLIPs.

Finally, we analyze the set of models derived by \textcite{nielsen2025} for the adsorption energies of formate on metal-supported metal oxide nanoclusters.
The derived models offer a range of physical interpretability, and include RBF-GPR, XGBoost, SISSO, and WWL-GPR.
The errors of these models are highlighted in \cref{supp/fig:nanoclusters-formate-adsorption-energy,supp/fig:nanoclusters-formate-adsorption-energy-relaxed} for the comparison between evaluations on the DFT local minima and on the MLIP local minima, respectively (see \cref{sec:results/relaxation}).
Differences between the task-specific models are discussed in more detail by \textcite{nielsen2025}, so they are not discussed further here.
Focusing first on the comparison with DFT local minima (\cref{supp/fig:nanoclusters-formate-adsorption-energy}), we see the cluster of best-performing MLIPs outperforming the task-specific models, achieving up to twice as good RMSE performance.
However, as discussed in \cref{sec:results/relaxation}, the best-performing MLIPs are most strongly affected by MLIP relaxation.
Hence, the picture changes when focusing on the comparison with MLIP local minima (\cref{supp/fig:nanoclusters-formate-adsorption-energy-relaxed}).
In this case, the task-specific models are much more performant compared to the MLIPs; the best-performing models even outperforming the best MLIPs.

In general, the task-specific models discussed above outperform the foundational MLIPs in all cases, although the differences are quite slim in some cases, and in the final example the models only outperformed MLIPs when comparing against MLIP local minima.

It is worth highlighting that the task-specific models predict properties for relaxed geometries using feature inputs only determined from non-relaxed geometries---the clean surfaces and information regarding the adsorption motif.
Therefore, applying these models is significantly cheaper and easier than applying a foundational MLIP for obtaining a property value, as the latter requires creating an adsorption geometry and relaxing this geometry.
However, it should also be reiterated that the task-specific models have limited applicability, and the nature of their construction means they cannot be compared one-to-one to foundational MLIPs.
This is discussed in more detail in the next section.

\section{Discussion}

In this work, we have systematically evaluated zero-shot performance of \num{80} different pretrained MLIPs on \num{13} properties relevant to heterogeneous catalysis, across \num{6} different data sets, obtaining over \num{700000} potential energy evaluations.
Our findings show that there is a large spread of accuracy between the different MLIPs, but more importantly, that there is no singular MLIP that is universally the best-performing model.
The best-performing MLIP differs between the different data sets evaluated in this work, but also differs between target properties within the same data set (i.e., reaction energy and adsorbate formation energy).
It is therefore imperative that one carefully investigates which foundational MLIP is most suitable for the desired application, as there is no generically applicable answer to this question.
Based on the findings in this work, the eSEN, Orb, and UMA models seem to be among the best-performing models across most of the evaluated properties.
Therefore, it is recommended to use these as a starting point for exploring model performance.

For the properties based on an energy difference (all except for zero-point energy), we observe that the best models can be very accurate, with RMSEs generally on the order of \qtyrange{0.1}{0.3}{\electronvolt} and maximum errors on the order of \qtyrange{0.2}{2}{\electronvolt}.
In this comparison, it is important to keep in mind that both the DFT data sets that the foundational MLIPs have been trained on and the catalysis-related data sets assembled in this work have been obtained with different DFT functionals and numerical settings.
Furthermore, depending on the property of interest, commonly used DFT functionals may have a sizable intrinsic error compared to experiments.
For instance, one study focusing on surface reaction energies found that functionals such as PBE, RPBE and BEEF-vdW may exhibit errors around \qtyrange{0.2}{0.3}{\electronvolt} or larger relative to experiment~\cite{wellendorff2015}.
Nevertheless, we are not yet at a point where the additional error introduced by replacing DFT with a foundational MLIP is negligible compared to the intrinsic DFT error.
Therefore, caution is suggested when using MLIP results for further analysis (i.e., microkinetic models based on activation energies).
However, the current-generation MLIPs are more than accurate enough for initial screening purposes, i.e., to approximate a desired property value, narrowing down a large amount of candidates into a smaller set, which can then be evaluated more accurately with suitable DFT settings.
Furthermore, it can be expected that MLIP accuracy will continue to increase in the coming years.

In addition to comparing foundational MLIPs to each other, we also compared them to a range of task-specific models trained on the specific data sets explored in this work.
These models include simple linear scaling relations, more advanced graph-based GPR models, and decision tree models.
We show that these models can compete with the current generation of best-performing MLIPs.
However, it is important to note that differences in their construction to that of MLIPs mean that they are not equally easy to apply.
Crucially, training task-specific models requires knowing DFT energies for relevant geometries \emph{a priori}, which is usually time-consuming and, in case of for example TS determination, nontrivial.
The task-specific models explored in this work have been trained on a large subset of the full data sets, whereas the foundational MLIPs have been trained on none of the geometries evaluated in this work.
It is therefore impressive that current-generation foundational MLIPs can obtain accuracy to this order of magnitude.
They achieve this due to their immense size---and the immense size of the training data sets---which is also a disadvantage, as inference times of MLIPs are much larger than those of the much simpler task-specific models.
Furthermore, some of the task-specific models (i.e., XGBoost and SISSO) offer model interpretability, providing insight in the underlying structure--property correlations~\cite{nielsen2025}.
The foundational MLIPs, by virtue of being immensely large graph neural networks, are entirely black-box models and can offer none of this interpretability.

One area of foundational MLIP application unexplored in this work, which can bridge the gap between DFT and MLIP performance described above, is augmentation of pretrained MLIPs, such as fine-tuning or Δ-learning.
In fine-tuning, training of an MLIP is continued with a different data set, which more closely approximates the target application of the MLIP.
Fine-tuning is already applied as part of the training protocol of many foundational MLIPs such as Orb~\cite{neumann2024,rhodes2025}, eSEN~\cite{fu2025}, and UMA~\cite{wood2025}, where the large training data sets discussed in \cref{sec:methods/models} are utilized.
However, fine-tuning can also be applied downstream using a small data set, with the goal of obtaining a model that is more accurate specifically in the region of this data set, and where loss of general performance is acceptable.
This type of fine-tuning has been successfully applied to a wide range of problems, from more general problems such as correcting for softening in MLIPs~\cite{deng2024} to more specific materials science applications~\cite{kaur2025,pitfield2025a,dellapia2025,radova2025,liu2025}.
In the Δ-learning approach~\cite{ramakrishnan2015}, a machine learning model is trained with the goal to accurately predict the error between a baseline calculation and the target calculation, such that the output of this model can be added to the baseline calculation to more closely approximate the target value.
In the context of MLIPs, the baseline calculation is the foundational MLIP prediction.
Here, Δ-learning has been successfully applied to a range of different materials science applications~\cite{lyngby2024,pitfield2025a,pitfield2025b}, although it has also been shown to not always be competitive with training an MLIP from scratch~\cite{radova2025}.
Overall, augmentation of foundational MLIPs can be a powerful method of training highly accurate, but less general, custom potentials for specific applications.
However, the approaches discussed above require initial collection of a training data set suitable for the desired application, which introduces similar disadvantages to those observed with the use of task-specific models.
Additionally, the question of which data to include in such a training data set---both in terms of data set size as well as coverage of the parameter space---is nontrivial.
Furthermore, many hyperparameters and workflow choices are involved in setting up a fine-tuning or Δ-learning workflow, which can all impact the quality of the resulting model.
As pretrained MLIPs generally consist of millions of parameters, the computational resources required to fine-tune a pretrained MLIP can also be high, even if the fine-tuning data set is relatively small---especially when some sort of hyperparameter optimization is additionally performed.
In short, the above reasons mean zero-shot application of a pretrained MLIP is much more desirable than augmentation, although the current-generation pretrained MLIPs are not quite yet at the accuracy required for this to be fully feasible across all applications.

\section*{Supplementary material}

Overview of evaluated MLIPs, model training performance (figures and tables) for all properties, and additional figures on poor prediction of magnetic elements.

\section*{Acknowledgments}

The authors would like to acknowledge funding from VILLUM FONDEN (grant no.\@ 37381), Novo Nordisk Fonden (grant no.\@ NNF22OC0078939) and the Danish National Research Foundation through the Center of Excellence `InterCat' (grant no.\@ DNRF150). Computational support was provided by the Centre for Scientific Computing Aarhus (CSCAA) at Aarhus University.

\section*{Author declarations}

\subsection*{Conflict of interest}

The authors have no conflicts to disclose.

\subsection*{Author contributions}

\textbf{Luuk H. E. Kempen}: Conceptualization (equal); Investigation (lead); Methodology (lead); Software (lead); Visualization (lead), Writing -- original draft (lead); Writing -- review \& editing (equal).
\textbf{Raffaele Cheula}: Investigation (supporting); Visualization (supporting); Writing -- review \& editing (equal).
\textbf{Mie Andersen}: Conceptualization (equal); Investigation (supporting); Writing -- review \& editing (equal); Supervision (lead); Resources (lead); Funding acquisition (lead).

\section*{Data availability}

The data that support the findings of this study will be made available on Zenodo upon publication under a CC BY 4.0 license.

\section*{Code availability}

The source code will be made available on GitLab upon publication under a GNU GPLv3 license.

\printbibliography

\clearpage

\onecolumn
\raggedbottom

\setcounter{page}{1}
\setcounter{section}{0}
\setcounter{table}{0}
\setcounter{figure}{0}
\setcounter{equation}{0}
\renewcommand{\thepage}{S\arabic{page}}
\renewcommand{\thesection}{S\arabic{section}}
\renewcommand{\thetable}{S\arabic{table}}
\renewcommand{\thefigure}{S\arabic{figure}}
\renewcommand{\theequation}{S\arabic{equation}}

\makeatletter
\let\oldtitle\@title
\title{Supplementary Information for ``\oldtitle''}
\date{}
\makeatother

\maketitle

\begin{refsection}

\section{Machine learning interatomic potentials}

\newcommand*{\rot}[1]{\rotatebox[origin=rB]{270}{#1}}
\newcommand*{\fnote}[1]{\makebox[0pt][l]{\textsuperscript{#1}}}

\begin{longtable}{l@{}l@{}l*{13}c}
\caption{An overview of all MLIPs evaluated in this study, organized by model family and training data set. Every bullet corresponds to an evaluated MLIP. \label{supp/tab:mlips}} \\
\toprule
&&& \multicolumn{13}{c}{Training data set} \\
\cmidrule(l){4-16}
Model family && Variant &
\rot{MPTrj} &
\rot{MPTrj\,+\,(s)Alex} &
\rot{OMat24} &
\rot{OMat24\,\textrightarrow\,MPTrj\,+\,sAlex} &
\rot{MatPES-PBE} &
\rot{OC20} &
\rot{OC22} &
\rot{OC20\,+\,OC22} &
\rot{OC20\,\textrightarrow\,OC22} &
\rot{OC25} &
\rot{OC20\,+\,AQCat25} &
\rot{OC20\,\textrightarrow\,AQCat25} &
\rot{OC20\,\textrightarrow\,OC20\,+\,AQCat25} \\
\midrule
\endhead
\bottomrule \addlinespace[\belowrulesep]
\caption*{\textsuperscript{a}~No strict neighbor enforcement; \textsuperscript{b}~subsampled Alexandria; \textsuperscript{c}~full Alexandria; \textsuperscript{d}~only AIMD subset; \textsuperscript{e}~All split; \textsuperscript{f}~All+MD split; \textsuperscript{g}~2M split; \textsuperscript{h}~extended (OC20++).}
\endfoot
Allegro && L-0.1 &\textbullet&&&\textbullet&&&&&&&&&\\
\multirow{4}{*}{AQCat} & \ldelim\{{4}{4mm} & 31M &&&&&&&&&&&&\textbullet&\\ \nopagebreak
&& 153M &&&&&&&&&&&&\textbullet&\\ \nopagebreak
&& 31M inFiLM &&&&&&&&&&&&&\textbullet\\ \nopagebreak
&& 31M in+midFiLM &&&&&&&&&&&\textbullet&&\\
CHGNet &&  &\textbullet&&&&&&&&&&&&\\
DimeNet\textsuperscript{++} &&  &&&&&&\textbullet\fnote{e}&&&&&&&\\
\multirow{8}{*}{EquiformerV2} & \ldelim\{{8}{4mm} & 31M &\textbullet&&\textbullet&\textbullet&&\textbullet\fnote{f}&&&&&&&\\
\end{longtable}

\clearpage

\section{Model prediction performance}  \label{supp/sec:model-performance}

\begin{figure}[H]
    \centering
    \includegraphics{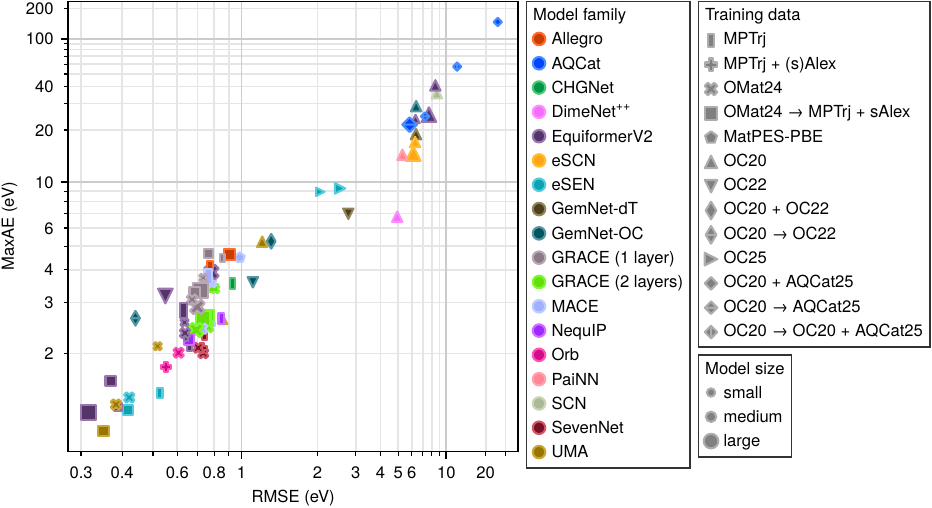}
    \caption{Prediction performance of the MLIPs for the slab vacancy formation energy of perovskite oxide surfaces.}
    \label{supp/fig:perovskites-slab-vacancy-formation-energy}
\end{figure}

\vspace{-1.5em}

\begin{longtable}{l@{}l@{}l@{}l@{}lS[table-format=3.3]S[table-format=3.3]}
\caption{Prediction performance of the MLIPs for the slab vacancy formation energy of perovskite oxide surfaces. \label{supp/tab:perovskitesdataset-slab_vacancy_formation_energy}} \\
\toprule
Model family && Variant && Training data set & {RMSE (\unit{\electronvolt})} & {MaxAE (\unit{\electronvolt})} \\
\midrule
\endhead
\bottomrule \addlinespace[\belowrulesep]
\caption*{\textsuperscript{a}~No strict neighbor enforcement.}
\endfoot
\multirow{2}{*}{Allegro} && \multirow{2}{*}{L-0.1} & \ldelim\{{2}{4mm} & MPTrj & 0.774 & 4.13 \\ \nopagebreak
&& && OMat24\,\textrightarrow\,MPTrj\,+\,sAlex & 0.904 & 4.62 \\
\multirow{4}{*}{AQCat} &\ldelim\{{4}{4mm} & 31M &&OC20\,\textrightarrow\,AQCat25 & 7.25 & 24.5 \\ \nopagebreak
&& 153M &&OC20\,\textrightarrow\,AQCat25 & 5.82 & 21.6 \\ \nopagebreak
&& 31M inFiLM &&OC20\,\textrightarrow\,OC20\,+\,AQCat25 & 12.0 & 57.3 \\ \nopagebreak
&& 31M in+midFiLM &&OC20\,+\,AQCat25 & 25.4 & 145. \\
CHGNet && &&MPTrj & 0.927 & 3.53 \\
DimeNet\textsuperscript{++} && &&OC20 (All) & 4.94 & 6.77 \\
\end{longtable}

\clearpage

\begin{figure}[H]
    \centering
    \includegraphics{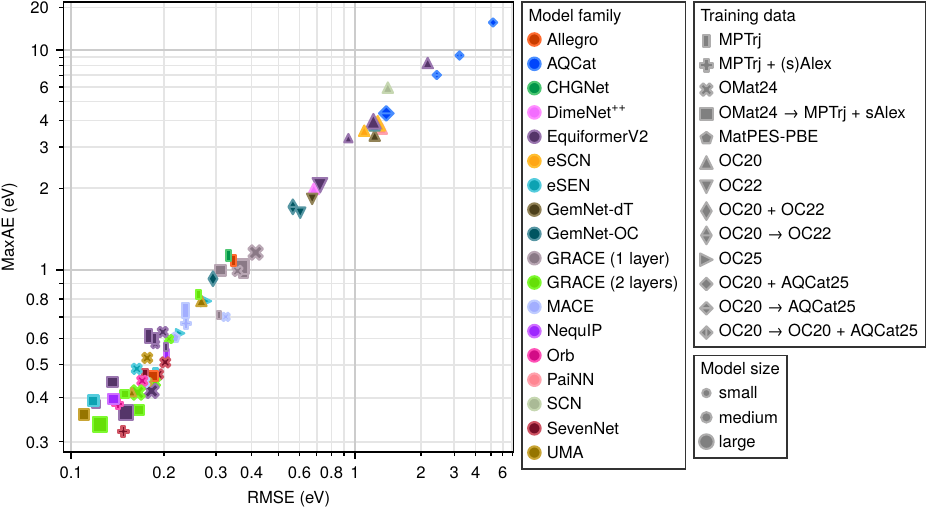}
    \caption{Prediction performance of the MLIPs for the activation energy of \ce{CO2} hydrogenation reaction steps on (doped) \ce{ZrO2(101)} surfaces.}
    \label{supp/fig:zirconia-activation-energy}
\end{figure}

\vspace{-1.5em}

\begin{longtable}{l@{}l@{}l@{}l@{}lS[table-format=2.3]S[table-format=2.3]}
\caption{Prediction performance of the MLIPs for the activation energy of (doped) \ce{ZrO2(101)} surfaces. \label{supp/tab:zirconiadataset-activation_energy}} \\
\toprule
Model family && Variant && Training data set & {RMSE (\unit{\electronvolt})} & {MaxAE (\unit{\electronvolt})} \\
\midrule
\endhead
\bottomrule \addlinespace[\belowrulesep]
\caption*{\textsuperscript{a}~No strict neighbor enforcement.}
\endfoot
\multirow{2}{*}{Allegro} && \multirow{2}{*}{L-0.1} & \ldelim\{{2}{4mm} & MPTrj & 0.346 & 1.08 \\ \nopagebreak
&& && OMat24\,\textrightarrow\,MPTrj\,+\,sAlex & 0.183 & 0.463 \\
\multirow{4}{*}{AQCat} &\ldelim\{{4}{4mm} & 31M &&OC20\,\textrightarrow\,AQCat25 & 2.41 & 7.02 \\ \nopagebreak
&& 153M &&OC20\,\textrightarrow\,AQCat25 & 1.37 & 4.35 \\ \nopagebreak
&& 31M inFiLM &&OC20\,\textrightarrow\,OC20\,+\,AQCat25 & 3.21 & 9.22 \\ \nopagebreak
&& 31M in+midFiLM &&OC20\,+\,AQCat25 & 5.16 & 15.4 \\
CHGNet && &&MPTrj & 0.331 & 1.12 \\
DimeNet\textsuperscript{++} && &&OC20 (All) & 0.681 & 1.99 \\
\end{longtable}

\clearpage

\begin{figure}[H]
    \centering
    \includegraphics{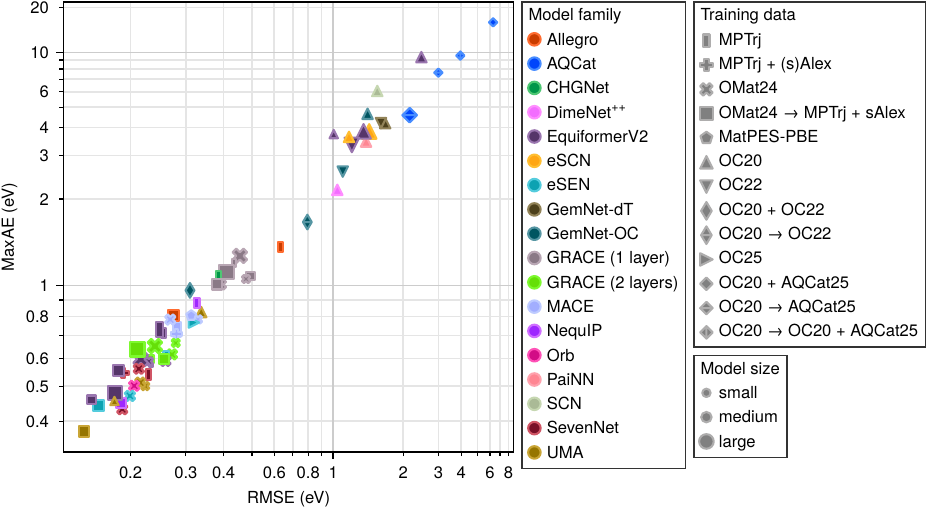}
    \caption{Prediction performance of the MLIPs for the reaction energy of \ce{CO2} hydrogenation reaction steps on (doped) \ce{ZrO2(101)} surfaces.}
    \label{supp/fig:zirconia-reaction-energy}
\end{figure}

\vspace{-1.5em}

\begin{longtable}{l@{}l@{}l@{}l@{}lS[table-format=2.3]S[table-format=2.3]}
\caption{Prediction performance of the MLIPs for the reaction energy of (doped) \ce{ZrO2(101)} surfaces. \label{supp/tab:zirconiadataset-reaction_energy}} \\
\toprule
Model family && Variant && Training data set & {RMSE (\unit{\electronvolt})} & {MaxAE (\unit{\electronvolt})} \\
\midrule
\endhead
\bottomrule \addlinespace[\belowrulesep]
\caption*{\textsuperscript{a}~No strict neighbor enforcement.}
\endfoot
\multirow{2}{*}{Allegro} && \multirow{2}{*}{L-0.1} & \ldelim\{{2}{4mm} & MPTrj & 0.632 & 1.34 \\ \nopagebreak
&& && OMat24\,\textrightarrow\,MPTrj\,+\,sAlex & 0.274 & 0.805 \\
\multirow{4}{*}{AQCat} &\ldelim\{{4}{4mm} & 31M &&OC20\,\textrightarrow\,AQCat25 & 2.99 & 7.61 \\ \nopagebreak
&& 153M &&OC20\,\textrightarrow\,AQCat25 & 2.15 & 4.57 \\ \nopagebreak
&& 31M inFiLM &&OC20\,\textrightarrow\,OC20\,+\,AQCat25 & 3.95 & 9.57 \\ \nopagebreak
&& 31M in+midFiLM &&OC20\,+\,AQCat25 & 6.28 & 15.7 \\
CHGNet && &&MPTrj & 0.386 & 1.07 \\
DimeNet\textsuperscript{++} && &&OC20 (All) & 1.04 & 2.17 \\
\end{longtable}

\clearpage

\begin{figure}[H]
    \centering
    \includegraphics{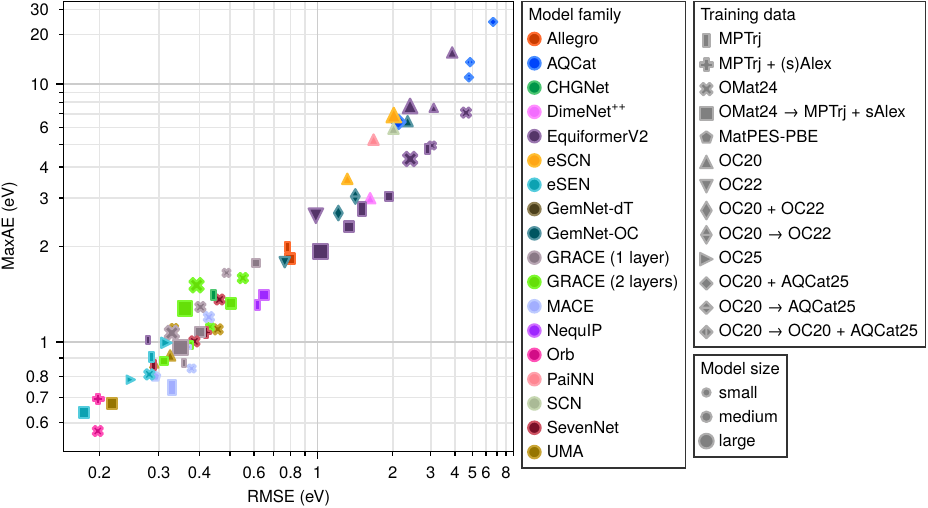}
    \caption{Prediction performance of the MLIPs for the formation energy of adsorbates of \ce{CO2} hydrogenation reaction steps on (doped) \ce{ZrO2(101)} surfaces.}
    \label{supp/fig:zirconia-formation-energy-ads}
\end{figure}

\vspace{-1.5em}

\begin{longtable}{l@{}l@{}l@{}l@{}lS[table-format=2.3]S[table-format=2.3]}
\caption{Prediction performance of the MLIPs for the formation energy of adsorbates of (doped) \ce{ZrO2(101)} surfaces. \label{supp/tab:zirconiadataset-formation_energy_ads}} \\
\toprule
Model family && Variant && Training data set & {RMSE (\unit{\electronvolt})} & {MaxAE (\unit{\electronvolt})} \\
\midrule
\endhead
\bottomrule \addlinespace[\belowrulesep]
\caption*{\textsuperscript{a}~No strict neighbor enforcement.}
\endfoot
\multirow{2}{*}{Allegro} && \multirow{2}{*}{L-0.1} & \ldelim\{{2}{4mm} & MPTrj & 0.779 & 1.99 \\ \nopagebreak
&& && OMat24\,\textrightarrow\,MPTrj\,+\,sAlex & 0.796 & 1.82 \\
\multirow{4}{*}{AQCat} &\ldelim\{{4}{4mm} & 31M &&OC20\,\textrightarrow\,AQCat25 & 4.77 & 10.9 \\ \nopagebreak
&& 153M &&OC20\,\textrightarrow\,AQCat25 & 2.12 & 6.38 \\ \nopagebreak
&& 31M inFiLM &&OC20\,\textrightarrow\,OC20\,+\,AQCat25 & 4.84 & 13.4 \\ \nopagebreak
&& 31M in+midFiLM &&OC20\,+\,AQCat25 & 6.63 & 24.4 \\
CHGNet && &&MPTrj & 0.442 & 1.38 \\
DimeNet\textsuperscript{++} && &&OC20 (All) & 1.60 & 2.99 \\
\end{longtable}

\clearpage

\begin{figure}[H]
    \centering
    \includegraphics{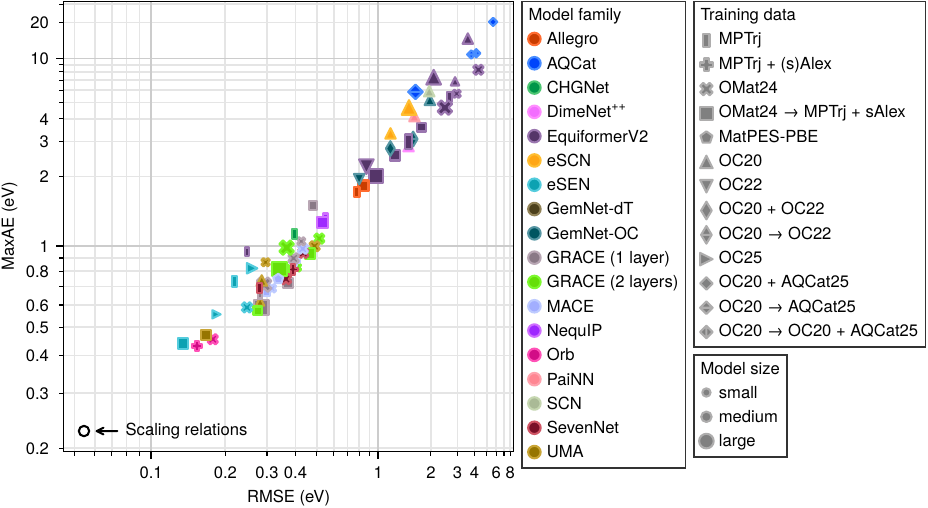}
    \caption{Prediction performance of the MLIPs for the formation energy of TSs of \ce{CO2} hydrogenation reaction steps on (doped) \ce{ZrO2(101)} surfaces.}
    \label{supp/fig:zirconia-formation-energy-ts}
\end{figure}

\vspace{-1.5em}

\begin{longtable}{l@{}l@{}l@{}l@{}lS[table-format=2.4]S[table-format=2.4]}
\caption{Prediction performance of the MLIPs for the formation energy of TSs of (doped) \ce{ZrO2(101)} surfaces. \label{supp/tab:zirconiadataset-formation_energy_ts}} \\
\toprule
Model family && Variant && Training data set & {RMSE (\unit{\electronvolt})} & {MaxAE (\unit{\electronvolt})} \\
\midrule
\endhead
\bottomrule \addlinespace[\belowrulesep]
\caption*{\textsuperscript{a}~No strict neighbor enforcement.}
\endfoot
\multirow{2}{*}{Allegro} && \multirow{2}{*}{L-0.1} & \ldelim\{{2}{4mm} & MPTrj & 0.778 & 1.70 \\ \nopagebreak
&& && OMat24\,\textrightarrow\,MPTrj\,+\,sAlex & 0.838 & 1.81 \\
\multirow{4}{*}{AQCat} &\ldelim\{{4}{4mm} & 31M &&OC20\,\textrightarrow\,AQCat25 & 3.78 & 10.8 \\ \nopagebreak
&& 153M &&OC20\,\textrightarrow\,AQCat25 & 1.61 & 5.85 \\ \nopagebreak
&& 31M inFiLM &&OC20\,\textrightarrow\,OC20\,+\,AQCat25 & 4.12 & 11.0 \\ \nopagebreak
&& 31M in+midFiLM &&OC20\,+\,AQCat25 & 5.63 & 20.2 \\
CHGNet && &&MPTrj & 0.396 & 1.12 \\
DimeNet\textsuperscript{++} && &&OC20 (All) & 1.47 & 2.83 \\
\end{longtable}

\clearpage

\begin{figure}[H]
    \centering
    \includegraphics{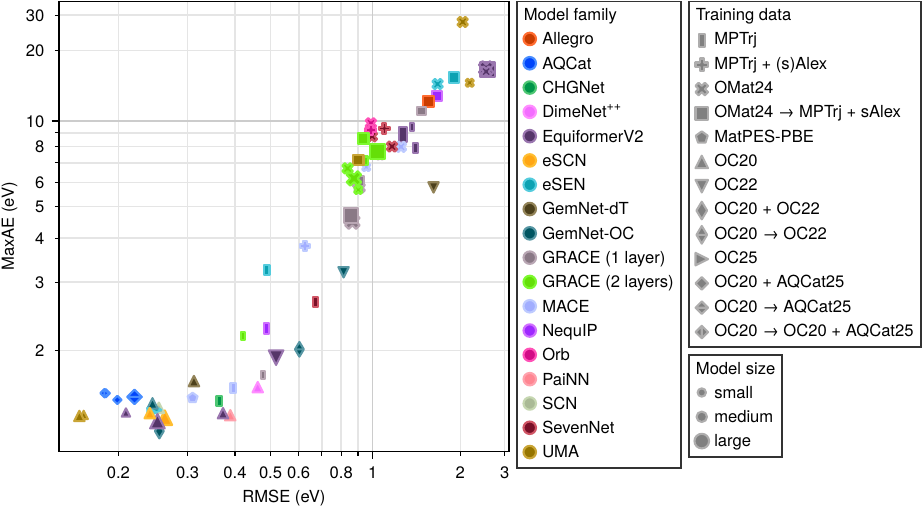}
    \caption{Prediction performance of the MLIPs for the activation energy of RWGS reaction steps on metals and SAAs.}
    \label{supp/fig:single-atom-alloys-activation-energy}
\end{figure}

\vspace{-1.5em}

\begin{longtable}{l@{}l@{}l@{}l@{}lS[table-format=2.3]S[table-format=2.3]}
\caption{Prediction performance of the MLIPs for the activation energy of RWGS reaction steps on metals and SAAs. \label{supp/tab:singleatomalloysdataset-activation_energy}} \\
\toprule
Model family && Variant && Training data set & {RMSE (\unit{\electronvolt})} & {MaxAE (\unit{\electronvolt})} \\
\midrule
\endhead
\bottomrule \addlinespace[\belowrulesep]
\caption*{\textsuperscript{a}~No strict neighbor enforcement.}
\endfoot
\multirow{2}{*}{Allegro} && \multirow{2}{*}{L-0.1} & \ldelim\{{2}{4mm} & MPTrj & 0.882 & 4.73 \\ \nopagebreak
&& && OMat24\,\textrightarrow\,MPTrj\,+\,sAlex & 1.53 & 12.0 \\
\multirow{4}{*}{AQCat} &\ldelim\{{4}{4mm} & 31M &&OC20\,\textrightarrow\,AQCat25 & 0.186 & 1.59 \\ \nopagebreak
&& 153M &&OC20\,\textrightarrow\,AQCat25 & 0.220 & 1.56 \\ \nopagebreak
&& 31M inFiLM &&OC20\,\textrightarrow\,OC20\,+\,AQCat25 & 0.185 & 1.59 \\ \nopagebreak
&& 31M in+midFiLM &&OC20\,+\,AQCat25 & 0.199 & 1.53 \\
CHGNet && &&MPTrj & 0.363 & 1.53 \\
DimeNet\textsuperscript{++} && &&OC20 (All) & 0.461 & 1.64 \\
\end{longtable}

\clearpage

\begin{figure}[H]
    \centering
    \includegraphics{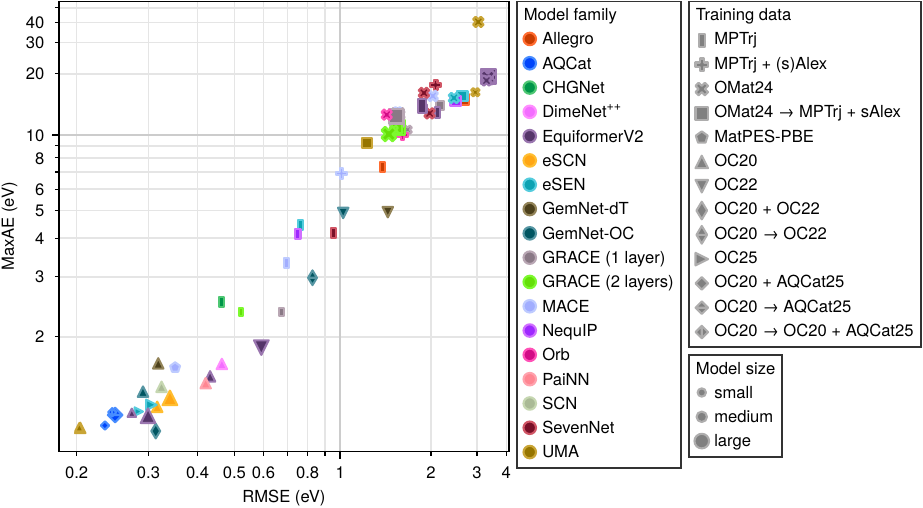}
    \caption{Prediction performance of the MLIPs for the reaction energy of RWGS reaction steps on metals and SAAs.}
    \label{supp/fig:single-atom-alloys-reaction-energy}
\end{figure}

\vspace{-1.5em}

\begin{longtable}{l@{}l@{}l@{}l@{}lS[table-format=2.3]S[table-format=2.3]}
\caption{Prediction performance of the MLIPs for the reaction energy of RWGS reaction steps on metals and SAAs. \label{supp/tab:singleatomalloysdataset-reaction_energy}} \\
\toprule
Model family && Variant && Training data set & {RMSE (\unit{\electronvolt})} & {MaxAE (\unit{\electronvolt})} \\
\midrule
\endhead
\bottomrule \addlinespace[\belowrulesep]
\caption*{\textsuperscript{a}~No strict neighbor enforcement.}
\endfoot
\multirow{2}{*}{Allegro} && \multirow{2}{*}{L-0.1} & \ldelim\{{2}{4mm} & MPTrj & 1.36 & 7.34 \\ \nopagebreak
&& && OMat24\,\textrightarrow\,MPTrj\,+\,sAlex & 2.65 & 14.7 \\
\multirow{4}{*}{AQCat} &\ldelim\{{4}{4mm} & 31M &&OC20\,\textrightarrow\,AQCat25 & 0.249 & 1.24 \\ \nopagebreak
&& 153M &&OC20\,\textrightarrow\,AQCat25 & 0.248 & 1.25 \\ \nopagebreak
&& 31M inFiLM &&OC20\,\textrightarrow\,OC20\,+\,AQCat25 & 0.244 & 1.27 \\ \nopagebreak
&& 31M in+midFiLM &&OC20\,+\,AQCat25 & 0.234 & 1.18 \\
CHGNet && &&MPTrj & 0.462 & 2.51 \\
DimeNet\textsuperscript{++} && &&OC20 (All) & 0.463 & 1.69 \\
\end{longtable}

\clearpage

\begin{figure}[H]
    \centering
    \includegraphics{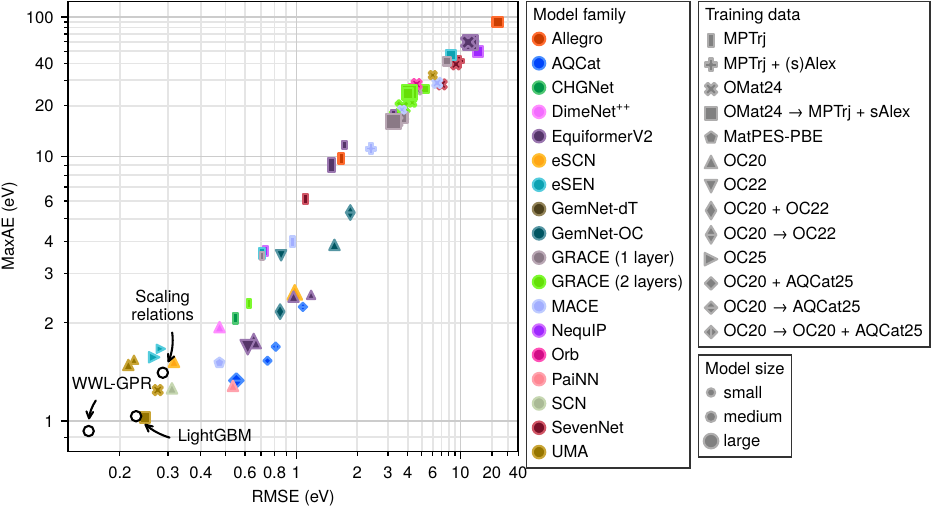}
    \caption{Prediction performance of the MLIPs for the formation energy of adsorbates of RWGS reaction steps on metals and SAAs.}
    \label{supp/fig:single-atom-alloys-formation-energy-ads}
\end{figure}

\vspace{2.5em}

\begin{longtable}{l@{}l@{}l@{}l@{}lS[table-format=2.3]S[table-format=2.3]}
\caption{Prediction performance of the MLIPs for the formation energy of adsorbates of RWGS reaction steps on metals and SAAs. \label{supp/tab:singleatomalloysdataset-formation_energy_ads}} \\
\toprule
Model family && Variant && Training data set & {RMSE (\unit{\electronvolt})} & {MaxAE (\unit{\electronvolt})} \\
\midrule
\endhead
\bottomrule \addlinespace[\belowrulesep]
\caption*{\textsuperscript{a}~No strict neighbor enforcement.}
\endfoot
\multirow{2}{*}{Allegro} && \multirow{2}{*}{L-0.1} & \ldelim\{{2}{4mm} & MPTrj & 1.65 & 9.77 \\ \nopagebreak
&& && OMat24\,\textrightarrow\,MPTrj\,+\,sAlex & 23.1 & 90.2 \\
\multirow{4}{*}{AQCat} &\ldelim\{{4}{4mm} & 31M &&OC20\,\textrightarrow\,AQCat25 & 1.07 & 2.27 \\ \nopagebreak
&& 153M &&OC20\,\textrightarrow\,AQCat25 & 0.553 & 1.31 \\ \nopagebreak
&& 31M inFiLM &&OC20\,\textrightarrow\,OC20\,+\,AQCat25 & 0.810 & 1.67 \\ \nopagebreak
&& 31M in+midFiLM &&OC20\,+\,AQCat25 & 0.746 & 1.51 \\
CHGNet && &&MPTrj & 0.550 & 2.07 \\
DimeNet\textsuperscript{++} && &&OC20 (All) & 0.474 & 1.94 \\
\end{longtable}

\clearpage

\begin{figure}[H]
    \centering
    \includegraphics{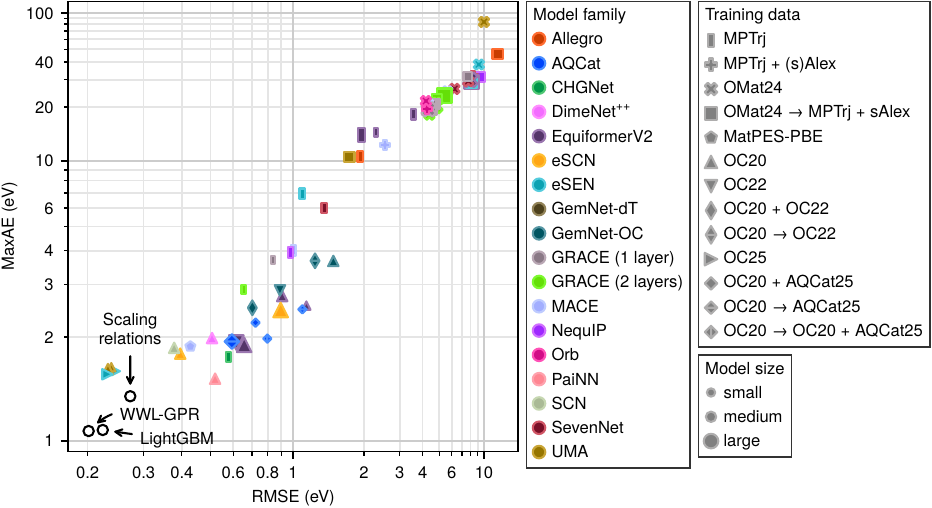}
    \caption{Prediction performance of the MLIPs for the formation energy of TSs of RWGS reaction steps on metals and SAAs.}
    \label{supp/fig:single-atom-alloys-formation-energy-ts}
\end{figure}

\vspace{2.5em}

\begin{longtable}{l@{}l@{}l@{}l@{}lS[table-format=2.3]S[table-format=2.3]}
\caption{Prediction performance of the MLIPs for the formation energy of TSs of RWGS reaction steps on metals and SAAs. \label{supp/tab:singleatomalloysdataset-formation_energy_ts}} \\
\toprule
Model family && Variant && Training data set & {RMSE (\unit{\electronvolt})} & {MaxAE (\unit{\electronvolt})} \\
\midrule
\endhead
\bottomrule \addlinespace[\belowrulesep]
\caption*{\textsuperscript{a}~No strict neighbor enforcement.}
\endfoot
\multirow{2}{*}{Allegro} && \multirow{2}{*}{L-0.1} & \ldelim\{{2}{4mm} & MPTrj & 1.92 & 10.5 \\ \nopagebreak
&& && OMat24\,\textrightarrow\,MPTrj\,+\,sAlex & 12.8 & 45.9 \\
\multirow{4}{*}{AQCat} &\ldelim\{{4}{4mm} & 31M &&OC20\,\textrightarrow\,AQCat25 & 1.09 & 2.46 \\ \nopagebreak
&& 153M &&OC20\,\textrightarrow\,AQCat25 & 0.593 & 1.93 \\ \nopagebreak
&& 31M inFiLM &&OC20\,\textrightarrow\,OC20\,+\,AQCat25 & 0.797 & 1.98 \\ \nopagebreak
&& 31M in+midFiLM &&OC20\,+\,AQCat25 & 0.721 & 2.22 \\
CHGNet && &&MPTrj & 0.579 & 1.73 \\
DimeNet\textsuperscript{++} && &&OC20 (All) & 0.506 & 1.98 \\
\end{longtable}

\clearpage

\begin{figure}[H]
    \centering
    \includegraphics{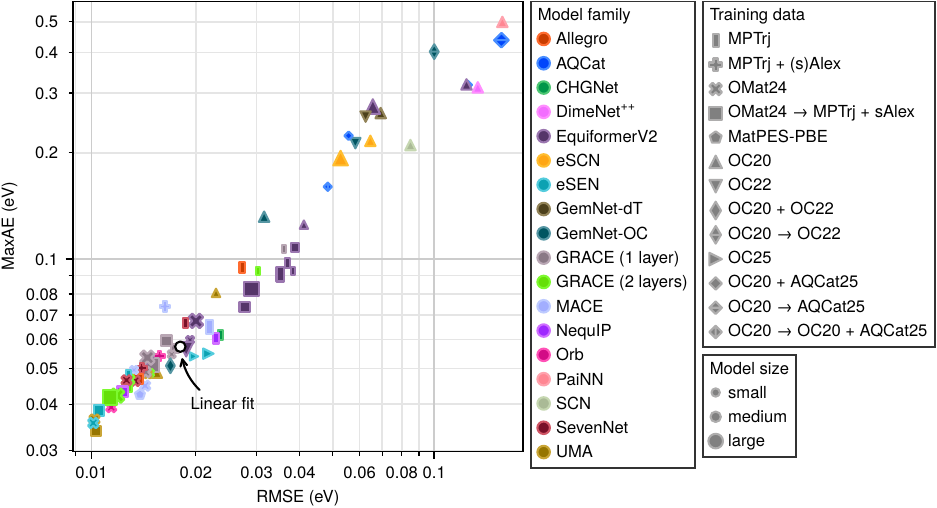}
    \caption{Prediction performance of the MLIPs for the zero-point energy of metal oxide nanoclusters on metal supports.}
    \label{supp/fig:nanoclusters-zero-point-energy}
\end{figure}

\vspace{-1.5em}

\begin{longtable}{l@{}l@{}l@{}l@{}lS[table-format=1.4]S[table-format=1.4]}
\caption{Prediction performance of the MLIPs for the zero-point energy of metal oxide nanoclusters on metal supports. \label{supp/tab:nanoclustersdataset-zero_point_energy}} \\
\toprule
Model family && Variant && Training data set & {RMSE (\unit{\electronvolt})} & {MaxAE (\unit{\electronvolt})} \\
\midrule
\endhead
\bottomrule \addlinespace[\belowrulesep]
\caption*{\textsuperscript{a}~No strict neighbor enforcement.}
\endfoot
\multirow{2}{*}{Allegro} && \multirow{2}{*}{L-0.1} & \ldelim\{{2}{4mm} & MPTrj & 0.0272 & 0.0947 \\ \nopagebreak
&& && OMat24\,\textrightarrow\,MPTrj\,+\,sAlex & 0.0136 & 0.0469 \\
\multirow{4}{*}{AQCat} &\ldelim\{{4}{4mm} & 31M &&OC20\,\textrightarrow\,AQCat25 & 0.126 & 0.318 \\ \nopagebreak
&& 153M &&OC20\,\textrightarrow\,AQCat25 & 0.160 & 0.437 \\ \nopagebreak
&& 31M inFiLM &&OC20\,\textrightarrow\,OC20\,+\,AQCat25 & 0.0484 & 0.160 \\ \nopagebreak
&& 31M in+midFiLM &&OC20\,+\,AQCat25 & 0.0557 & 0.224 \\
CHGNet && &&MPTrj & 0.0236 & 0.0617 \\
DimeNet\textsuperscript{++} && &&OC20 (All) & 0.135 & 0.312 \\
\end{longtable}

\clearpage

\begin{figure}[H]
    \centering
    \includegraphics{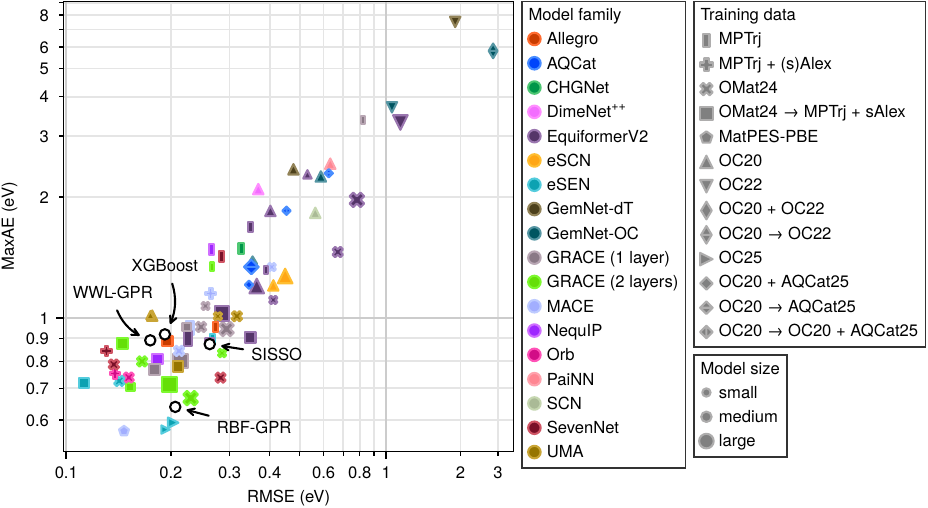}
    \caption{Prediction performance of the MLIPs for the adsorption energy of formate on metal-supported metal oxide nanoclusters.}
    \label{supp/fig:nanoclusters-formate-adsorption-energy}
\end{figure}

\vspace{2.5em}

\begin{longtable}{l@{}l@{}l@{}l@{}lS[table-format=1.3]S[table-format=1.3]}
\caption{Prediction performance of the MLIPs for the adsorption energy of formate on metal-supported metal oxide nanoclusters. \label{supp/tab:nanoclustersformatedataset-binding_energy}} \\
\toprule
Model family && Variant && Training data set & {RMSE (\unit{\electronvolt})} & {MaxAE (\unit{\electronvolt})} \\
\midrule
\endhead
\bottomrule \addlinespace[\belowrulesep]
\caption*{\textsuperscript{a}~No strict neighbor enforcement.}
\endfoot
\multirow{2}{*}{Allegro} && \multirow{2}{*}{L-0.1} & \ldelim\{{2}{4mm} & MPTrj & 0.272 & 0.953 \\ \nopagebreak
&& && OMat24\,\textrightarrow\,MPTrj\,+\,sAlex & 0.195 & 0.887 \\
\multirow{4}{*}{AQCat} &\ldelim\{{4}{4mm} & 31M &&OC20\,\textrightarrow\,AQCat25 & 0.625 & 2.33 \\ \nopagebreak
&& 153M &&OC20\,\textrightarrow\,AQCat25 & 0.350 & 1.32 \\ \nopagebreak
&& 31M inFiLM &&OC20\,\textrightarrow\,OC20\,+\,AQCat25 & 0.453 & 1.84 \\ \nopagebreak
&& 31M in+midFiLM &&OC20\,+\,AQCat25 & 0.344 & 1.20 \\
CHGNet && &&MPTrj & 0.325 & 1.47 \\
DimeNet\textsuperscript{++} && &&OC20 (All) & 0.368 & 2.10 \\
\end{longtable}

\clearpage

\begin{figure}[H]
    \centering
    \includegraphics{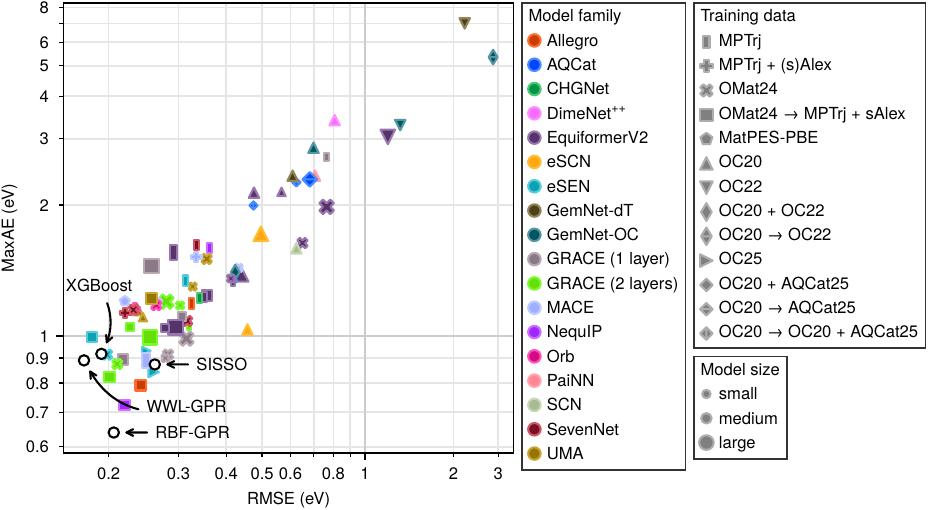}
    \caption{Prediction performance of the MLIPs for the adsorption energy of formate on metal-supported metal oxide nanoclusters, after relaxing into the local minimum for each MLIP.}
    \label{supp/fig:nanoclusters-formate-adsorption-energy-relaxed}
\end{figure}

\vspace{2.5em}

\begin{longtable}{l@{}l@{}l@{}l@{}lS[table-format=1.3]S[table-format=1.3]}
\caption{Prediction performance of the MLIPs for the adsorption energy of formate on metal-supported metal oxide nanoclusters, after relaxing into the local minimum for each MLIP. \label{supp/tab:nanoclustersformatedataset-binding_energy_relaxed}} \\
\toprule
Model family && Variant && Training data set & {RMSE (\unit{\electronvolt})} & {MaxAE (\unit{\electronvolt})} \\
\midrule
\endhead
\bottomrule \addlinespace[\belowrulesep]
\caption*{\textsuperscript{a}~No strict neighbor enforcement.}
\endfoot
\multirow{2}{*}{Allegro} && \multirow{2}{*}{L-0.1} & \ldelim\{{2}{4mm} & MPTrj & 0.325 & 1.18 \\ \nopagebreak
&& && OMat24\,\textrightarrow\,MPTrj\,+\,sAlex & 0.240 & 0.792 \\
\multirow{4}{*}{AQCat} &\ldelim\{{4}{4mm} & 31M &&OC20\,\textrightarrow\,AQCat25 & 0.624 & 2.29 \\ \nopagebreak
&& 153M &&OC20\,\textrightarrow\,AQCat25 & 0.682 & 2.33 \\ \nopagebreak
&& 31M inFiLM &&OC20\,\textrightarrow\,OC20\,+\,AQCat25 & 0.473 & 1.99 \\ \nopagebreak
&& 31M in+midFiLM &&OC20\,+\,AQCat25 & 0.417 & 1.32 \\
CHGNet && &&MPTrj & 0.340 & 1.21 \\
DimeNet\textsuperscript{++} && &&OC20 (All) & 0.807 & 3.39 \\
\end{longtable}

\clearpage

\begin{figure}[H]
    \centering
    \includegraphics{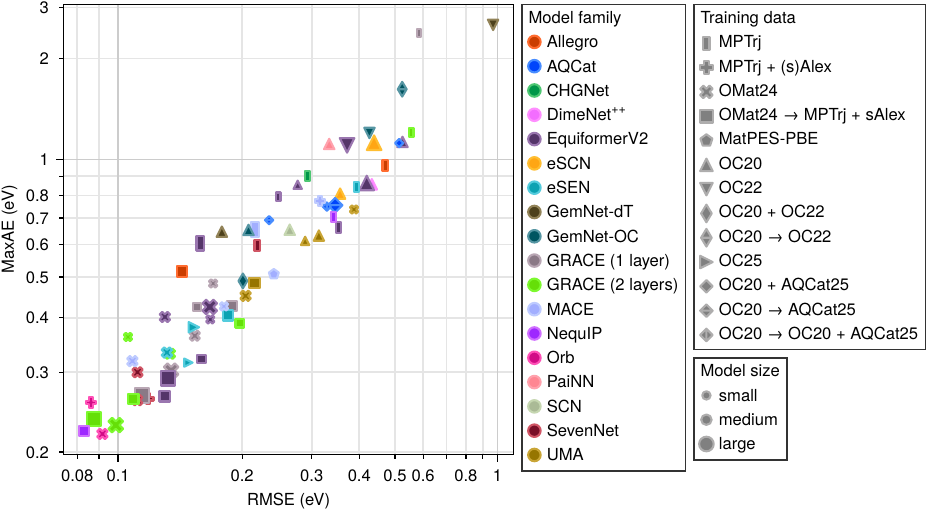}
    \caption{Prediction performance of the MLIPs for the activation energy of the \ce{CO2 + H <--> HCOO} reaction step on metal-supported metal oxide nanoclusters.}
    \label{supp/fig:nanoclusters-formate-reaction-activation-energy}
\end{figure}

\vspace{-1.5em}

\begin{longtable}{l@{}l@{}l@{}l@{}lS[table-format=1.4]S[table-format=1.4]}
\caption{Prediction performance of the MLIPs for the activation energy of the \ce{CO2 + H <--> HCOO} reaction step on metal-supported metal oxide nanoclusters. \label{supp/tab:nanoclustersformatereactiondataset-activation_energy}} \\
\toprule
Model family && Variant && Training data set & {RMSE (\unit{\electronvolt})} & {MaxAE (\unit{\electronvolt})} \\
\midrule
\endhead
\bottomrule \addlinespace[\belowrulesep]
\caption*{\textsuperscript{a}~No strict neighbor enforcement.}
\endfoot
\multirow{2}{*}{Allegro} && \multirow{2}{*}{L-0.1} & \ldelim\{{2}{4mm} & MPTrj & 0.470 & 0.964 \\ \nopagebreak
&& && OMat24\,\textrightarrow\,MPTrj\,+\,sAlex & 0.143 & 0.515 \\
\multirow{4}{*}{AQCat} &\ldelim\{{4}{4mm} & 31M &&OC20\,\textrightarrow\,AQCat25 & 0.234 & 0.691 \\ \nopagebreak
&& 153M &&OC20\,\textrightarrow\,AQCat25 & 0.345 & 0.755 \\ \nopagebreak
&& 31M inFiLM &&OC20\,\textrightarrow\,OC20\,+\,AQCat25 & 0.328 & 0.749 \\ \nopagebreak
&& 31M in+midFiLM &&OC20\,+\,AQCat25 & 0.514 & 1.11 \\
CHGNet && &&MPTrj & 0.292 & 0.901 \\
DimeNet\textsuperscript{++} && &&OC20 (All) & 0.432 & 0.857 \\
\end{longtable}

\clearpage

\begin{figure}[H]
    \centering
    \includegraphics{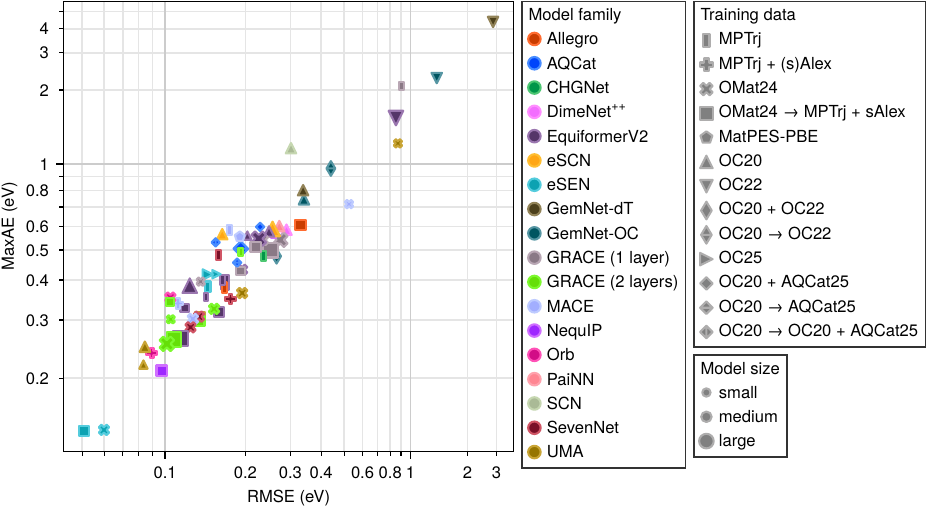}
    \caption{Prediction performance of the MLIPs for the reaction energy of the \ce{CO2 + H <--> HCOO} reaction step on metal-supported metal oxide nanoclusters.}
    \label{supp/fig:nanoclusters-formate-reaction-reaction-energy}
\end{figure}

\vspace{-1.5em}

\begin{longtable}{l@{}l@{}l@{}l@{}lS[table-format=1.4]S[table-format=1.4]}
\caption{Prediction performance of the MLIPs for the reaction energy of the \ce{CO2 + H <--> HCOO} reaction step on metal-supported metal oxide nanoclusters. \label{supp/tab:nanoclustersformatereactiondataset-reaction_energy}} \\
\toprule
Model family && Variant && Training data set & {RMSE (\unit{\electronvolt})} & {MaxAE (\unit{\electronvolt})} \\
\midrule
\endhead
\bottomrule \addlinespace[\belowrulesep]
\caption*{\textsuperscript{a}~No strict neighbor enforcement.}
\endfoot
\multirow{2}{*}{Allegro} && \multirow{2}{*}{L-0.1} & \ldelim\{{2}{4mm} & MPTrj & 0.166 & 0.378 \\ \nopagebreak
&& && OMat24\,\textrightarrow\,MPTrj\,+\,sAlex & 0.330 & 0.606 \\
\multirow{4}{*}{AQCat} &\ldelim\{{4}{4mm} & 31M &&OC20\,\textrightarrow\,AQCat25 & 0.186 & 0.455 \\ \nopagebreak
&& 153M &&OC20\,\textrightarrow\,AQCat25 & 0.192 & 0.505 \\ \nopagebreak
&& 31M inFiLM &&OC20\,\textrightarrow\,OC20\,+\,AQCat25 & 0.154 & 0.530 \\ \nopagebreak
&& 31M in+midFiLM &&OC20\,+\,AQCat25 & 0.228 & 0.598 \\
CHGNet && &&MPTrj & 0.235 & 0.478 \\
DimeNet\textsuperscript{++} && &&OC20 (All) & 0.289 & 0.584 \\
\end{longtable}

\clearpage

\section{Poor prediction of magnetic elements}

\begin{figure}[H]
    \centering
    \includegraphics{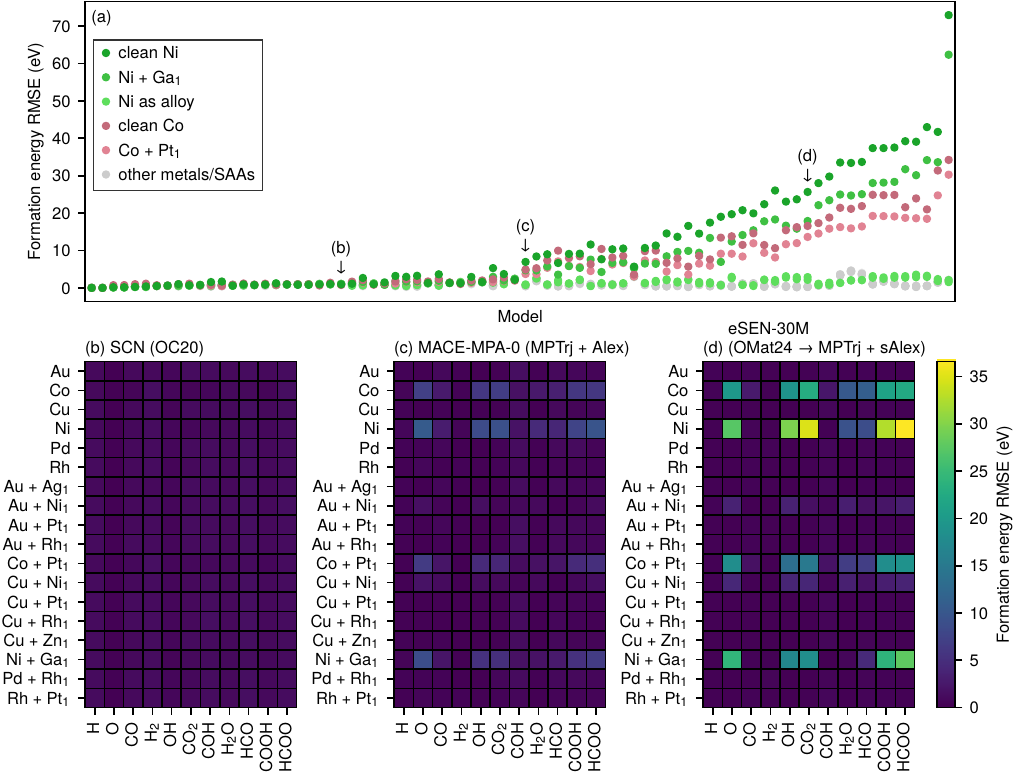}
    \caption{Formation energy predictions for reaction intermediates on metals and SAAs. (a) RMSE values for each combination of model and surface composition. Each column corresponds to one of the evaluated models, sorted by formation energy RMSE across the entire data set. Ni- and Co-containing surfaces have been indicated with green and red, respectively. (b--d) RMSE values for each combination of surface composition and adsorbate for three representative models: (b) a model trained on OC20, showing low errors for all combinations; (c) a model trained on MPTrj + Alex, showing somewhat increased errors for specific combinations; (d) a model trained on OAM, showing significant errors for specific combinations. Note that the errors in this figure have not been corrected as outlined in \cref{sec:methods/gas-phase-correction} as the high-error outliers would skew the results for the low-error combinations.}
    \label{supp/fig:saa-errors-ads}
\end{figure}

\begin{figure}[H]
    \centering
    \includegraphics{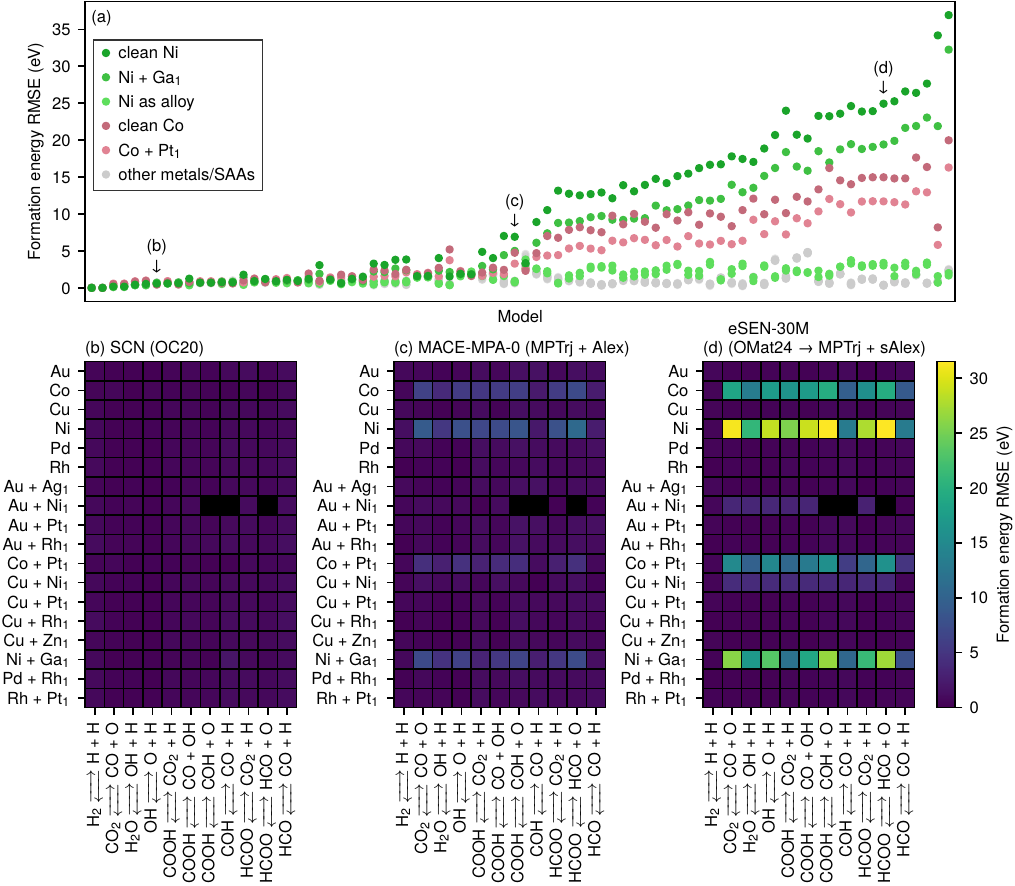}
    \caption{Formation energy predictions for TSs on metals and SAAs. (a) RMSE values for each combination of model and surface composition. Each column corresponds to one of the evaluated models, sorted by formation energy RMSE across the entire data set. Ni- and Co-containing surfaces have been indicated with green and red, respectively. (b--d) RMSE values for each combination of surface composition and reaction for three representative models: (b) a model trained on OC20, showing low errors for all combinations; (c) a model trained on MPTrj + Alex, showing somewhat increased errors for specific combinations; (d) a model trained on OAM, showing significant errors for specific combinations. Note that the errors in this figure have not been corrected as outlined in \cref{sec:methods/gas-phase-correction} as the high-error outliers would skew the results for the low-error combinations.}
    \label{supp/fig:saa-errors-ts}
\end{figure}

\clearpage

\printbibliography

\end{refsection}

\end{document}